\documentclass[11pt]{article}

\usepackage[a4paper,margin=1in]{geometry}
\usepackage{amsmath,amssymb,amsfonts,amsthm}
\usepackage{bm}
\usepackage{graphicx}
\usepackage{xparse}
\usepackage{subcaption}
\usepackage{booktabs}
\usepackage{array}
\usepackage{longtable}
\usepackage{multirow}
\usepackage{natbib}
\usepackage{xcolor}
\usepackage{url}

\usepackage{hyperref}
\hypersetup{
  colorlinks=true,
  linkcolor=blue,
  citecolor=blue,
  urlcolor=blue
}
\usepackage{pdflscape}

\usepackage{authblk}

\newcommand{\Var}{\operatorname{Var}}
\newcommand{\Cov}{\operatorname{Cov}}

\newcommand{\Prob}{\mathbb{P}}
\newcommand{\R}{\mathbb{R}}

\newcommand{\Normal}{\mathcal{N}}
\newcommand{\Bin}{\operatorname{Binomial}}

\newcommand{\diag}{\operatorname{diag}}

\newcommand{\rank}{\operatorname{rank}}

\newenvironment{keywords}{\par\smallskip\noindent\textbf{Keywords:}\ }{\par\smallskip}
\newenvironment{jelcodes}{\par\smallskip\noindent\textbf{JEL classification:}\ }{\par\smallskip}

\title{Temporal Coarse-Graining of Multi-Sector Default Count Data Generates Posterior-Implied Copulas}

\author{Shintaro Mori\thanks{Corresponding author. Email: shintaro.mori@gmail.com}}
\author[2]{Masato Hisakado}

\affil{Graduate School of Science and Technology, Hirosaki University}
\affil[2]{
College of Transdisciplinary Sciences for Innovation, Kanazawa university}

\begin{document}

\maketitle

\begin{abstract}
Sectoral default dependence is usually described by a static correlation matrix, a static copula, or a small number of common factors. Such representations, when specified separately at each observation horizon, do not by themselves explain why the effective dependence observed in monthly credit data differs from that observed after annual aggregation. This paper proposes a dynamic low-rank state-space model for monthly multi-sector default-count data and studies the dependence structure induced by temporal coarse-graining. The leading eigenvectors of the monthly sectoral default-rate correlation matrix are used as fixed loading directions for persistent AR(1) latent credit-state factors, and defaults are modeled through a binomial observation layer. Survival aggregation of monthly posterior probability paths induces horizon-dependent distributions of sectoral default-probability vectors, from which effective correlation matrices, eigenvalue spectra, and posterior-implied rank copulas are obtained. Applied to S\&P monthly sector-level default-count data from 1981--01 to 2021--09, a two-factor specification captures the dominant market-wide and sector-rotation modes, reproduces the annual amplification of the leading eigenvalues, and generates heterogeneous copula structures across sector pairs. 
In an annual forecast evaluation, the dynamic factor specifications reduce the
under-dispersion of static binomial and beta-binomial baselines, improving
interval coverage and CRPS for aggregate portfolio counts. In log-score-based
forecast comparisons, the one-factor specification is highly competitive,
whereas the two-factor specification improves sector-level calibration as
measured by per-sector CRPS.
\end{abstract}

\begin{keywords}
credit risk; default counts; dynamic factor model; temporal aggregation; copula;
state-space model; forecast evaluation
\end{keywords}

\begin{jelcodes}
C11; C32; C53; G21; G32; G33
\end{jelcodes}

\section{Introduction}
\label{sec:introduction}

Default clustering is a central issue in portfolio credit risk and
credit-derivatives modelling \citep{Schonbucher2003,McNeilFreyEmbrechts2015}.
A portfolio loss distribution depends not only on marginal default probabilities,
but also on how default risk co-moves across obligors, sectors, rating classes,
and time. Standard representations include static copulas, asset-correlation
models, and common-factor or frailty models
\citep{Vasicek1991,Vasicek2002,CreditMetrics1997,Li2000,
DasDuffieKapadiaSaita2007,DuffieEcknerHorelSaita2009}. Related approaches model clustering through
contagion, self-exciting count processes, or interacting default mechanisms
\citep{DavisLo2001,SakataHisakadoMori2007,TorriGiacomettiFarina2026,
Hawkes1971,ErraisGieseckeGoldberg2010,HisakadoHattoriMori2022RSS}.
These representations are useful for constructing portfolio loss distributions,
but they leave open a basic empirical question: why does the dependence
structure inferred from monthly default data differ from that inferred from
annual default data?

This question is particularly important when credit-risk data are temporally
aggregated. Annual default counts are informative for long-run risk management,
but they provide only a small number of observations and merge within-year
dynamics into a single count. In such aggregated data, different mechanisms can
be difficult to distinguish. Previous work on aggregated default counts showed
that contagion, common-factor dependence, and fluctuations in default
probabilities can generate similar long-horizon count distributions, making
mechanism identification difficult \citep{Mori2026ContagionMacro}. A subsequent
dynamic analysis showed that a persistent monthly latent default-probability
path can generate scale-dependent effective default correlation when
coarse-grained to longer horizons
\citep{Mori2026TemporalCoarseGraining}. These results suggest a conservative
baseline for the identification problem: part of the long-horizon dependence
that might otherwise be attributed to static correlation or contagion can be
generated by temporal persistence in the latent default-probability path alone.

The present paper extends this idea from a single aggregate default-count series
to a multi-sector credit portfolio. The object of interest is no longer a scalar
default probability or a scalar effective default correlation at a given
aggregation horizon, but a vector of sectoral default probabilities together with
horizon-dependent correlation matrices, posterior-implied copulas, and predictive
distributions of annual sectoral default counts. This extension matters for risk
management because a static correlation matrix or a static copula fitted at one
observation horizon does not specify how the sectoral dependence structure
changes when monthly credit states are temporally aggregated to annual
horizons. A model for sectoral default risk therefore needs not only a
cross-sectional dependence structure, but also a dynamic mechanism linking
monthly latent credit states to long-horizon portfolio and sector-vector loss
distributions.

The paper is also related to the literature on dynamic default prediction,
frailty-based default forecasting, and the empirical identification of default
clustering mechanisms. Multi-period corporate default prediction models
incorporate firm-level and macroeconomic covariates to estimate conditional
default probabilities over future horizons
\citep{DuffieSaitaWang2007}. Frailty and dynamic factor models further show
that latent common components remain important even after controlling for
observable macro-financial variables, and that such components can improve
portfolio-level default-risk measurement and forecasting
\citep{DuffieEcknerHorelSaita2009,KoopmanLucasSchwaab2011,
KoopmanLucasSchwaab2012,AzizpourGieseckeSchwenkler2018}. Closely related work
also emphasizes the difficulty of distinguishing contagion from conditional
independence or latent common risk in corporate default data
\citep{LandoNielsen2010}. The present paper does not aim to replace these
general default-prediction frameworks. Instead, it focuses on a complementary
question: how persistent low-rank sectoral credit-state dynamics estimated at
the monthly scale are transformed, through temporal aggregation, into
horizon-dependent sectoral correlation matrices, posterior-implied copulas, and
predictive default-count distributions.

The empirical starting point is the observation that sectoral default-rate
co-movements have a visible low-rank structure at the monthly scale. The first
empirical eigenvector of the monthly sectoral correlation matrix is naturally
interpreted as a market-wide default-risk mode, while the second captures a
sector-rotation mode that reallocates credit stress across sectors. A static
principal-component analysis can reveal these directions, but it does not specify
how monthly sectoral dynamics generate annual dependence.

The empirical starting point is the observation that sectoral default-rate
co-movements have a visible low-rank structure at the monthly scale. This is
consistent with empirical evidence from credit and CDS markets, where principal
component analyses often find that a dominant market-wide component explains a
large fraction of credit-spread or CDS co-movement
\citep{PanSingleton2008,AyadiBenAmeurGuesmi2013,MaldonadoQuintelaDelgado2020}.
The first empirical eigenvector of the monthly sectoral correlation matrix is
therefore naturally interpreted as a market-wide default-risk mode, while the
second captures a sector-rotation mode that reallocates credit stress across
sectors. A static principal-component analysis can reveal these directions, but
it does not specify how monthly sectoral dynamics generate annual dependence.

This paper therefore lifts these empirical eigenmodes into a Bayesian state-space
model. The first two empirical eigenvectors of the monthly sectoral correlation
matrix are used as fixed loading directions for persistent AR(1) latent factors,
and sectoral defaults are modeled through binomial observation equations. Monthly
posterior probability paths are then mapped to longer horizons by survival-based
temporal coarse-graining, producing horizon-dependent distributions of sectoral
default probabilities and annual default counts. Correlation matrices, eigenvalue
spectra, posterior-implied rank copulas, and predictive default-count
distributions are treated as summaries of these induced distributions. 
The
contribution is to show that this parsimonious monthly model reproduces the main
annual-scale dependence diagnostics and improves the calibration of annual
forecast distributions without fitting an independent annual model.

The remainder of this paper is organized as follows. Section~\ref{sec:data}
describes the data and the empirical evidence for temporal coarse-graining. It
also examines the empirical eigenmodes and motivates the two-factor
specification. Section~\ref{sec:model} defines the dynamic low-rank
AR(1)--Binomial model and the temporal coarse-graining map.
Section~\ref{sec:diagnostics} presents model diagnostics, including correlation
matrices, eigenvalue scaling, and posterior-implied copulas.
Section~\ref{sec:forecast} evaluates annual default-count forecasts.
Section~\ref{sec:discussion} discusses interpretation and limitations.
Section~\ref{sec:conclusion} concludes.

\section{Data and empirical coarse-graining diagnostics}
\label{sec:data}

\subsection{Monthly sector-level default-count panel}
\label{subsec:data_panel}

The empirical analysis uses monthly sector-level corporate default-count data
from Standard \& Poor's (S\&P). 
The sample period is January 1981 to September 2021, giving $T=489$ monthly
observations. Let $S=13$ denote the number of sectors. For each month $t=0,\ldots,T-1$ and sector $s=1,\ldots,S$, the cleaned panel
contains the number of obligors $N_{t,s}$, the number of defaults $L_{t,s}$, and
the default rate
\[
  r_{t,s}=\frac{L_{t,s}}{N_{t,s}}.
\]
The main panel consists of 13 sectoral series. An aggregate ALL series is
constructed by summing defaults and exposures across sectors. The temporal
coarse-graining of the ALL series and its consistency with annual default-count
benchmarks were examined in~\citet{Mori2026TemporalCoarseGraining}. In the
present paper, the ALL series is used only for descriptive checks and comparison
with the aggregate analysis; the main object of analysis is the 13-dimensional
sectoral default-count panel.

For an aggregation horizon $k$ measured
in months, non-overlapping block $b=0,1,\ldots$ consists of months
$t=bk,\ldots,bk+k-1$. The aggregated default count and beginning-of-block
exposure are defined by
\[
L_{b,s}^{(k)}=\sum_{j=0}^{k-1}L_{bk+j,s},
\qquad
N_{b,s}^{(k)}=N_{bk,s}.
\]
The monthly-equivalent $k$-month default rate is
\begin{equation}
r_{b,s}^{(k)}=\frac{1}{k}\frac{L_{b,s}^{(k)}}{N_{b,s}^{(k)}}.
\label{eq:monthly_equiv_rate}
\end{equation}
This normalization allows variance and correlation diagnostics to be compared
across horizons on a monthly scale. The use of beginning-of-block exposures is a
simple convention for constructing comparable non-overlapping panels, since
obligor counts can change within a block
\citep{Mori2026TemporalCoarseGraining}. Because sectoral monthly default
probabilities are small, this convention is adequate for the coarse-graining
diagnostics and is consistent with the survival-based probability aggregation
used in the model.

Table~\ref{tab:data_construction} summarizes the construction of the monthly
multi-sector default-count panel used throughout the empirical analysis. The
table reports the variables and aggregation conventions that enter the
coarse-graining diagnostics and the state-space model.

\begin{table}[htbp]
\centering
\caption{Construction of the monthly multi-sector default-count panel.}
\label{tab:data_construction}
\begin{tabular}{ll}
\toprule
Item & Description \\
\midrule
Data source & Standard \& Poor's monthly sector-level default-count data \\
Sample period & 1981--01 to 2021--09 \\
Number of months & $T=489$ \\
Main panel & 13 sectors \\
Additional series & Aggregate ALL series constructed from the 13 sectors \\
Observed variables & Obligors $N_{t,s}$, defaults $L_{t,s}$, rates $r_{t,s}$ \\
Aggregation horizons & $k=1,2,3,4,6,12$ months \\
Block exposure & Beginning-of-block exposure $N_{b,s}^{(k)}=N_{bk,s}$ \\
Monthly-equivalent rate & $r_{b,s}^{(k)}=L_{b,s}^{(k)}/(kN_{b,s}^{(k)})$ \\
\bottomrule
\end{tabular}
\end{table}

\subsection{Empirical variance and correlation scaling}

\begin{figure}[htbp]
\centering
\begin{subfigure}[t]{0.48\textwidth}
\centering
\includegraphics[width=\textwidth]{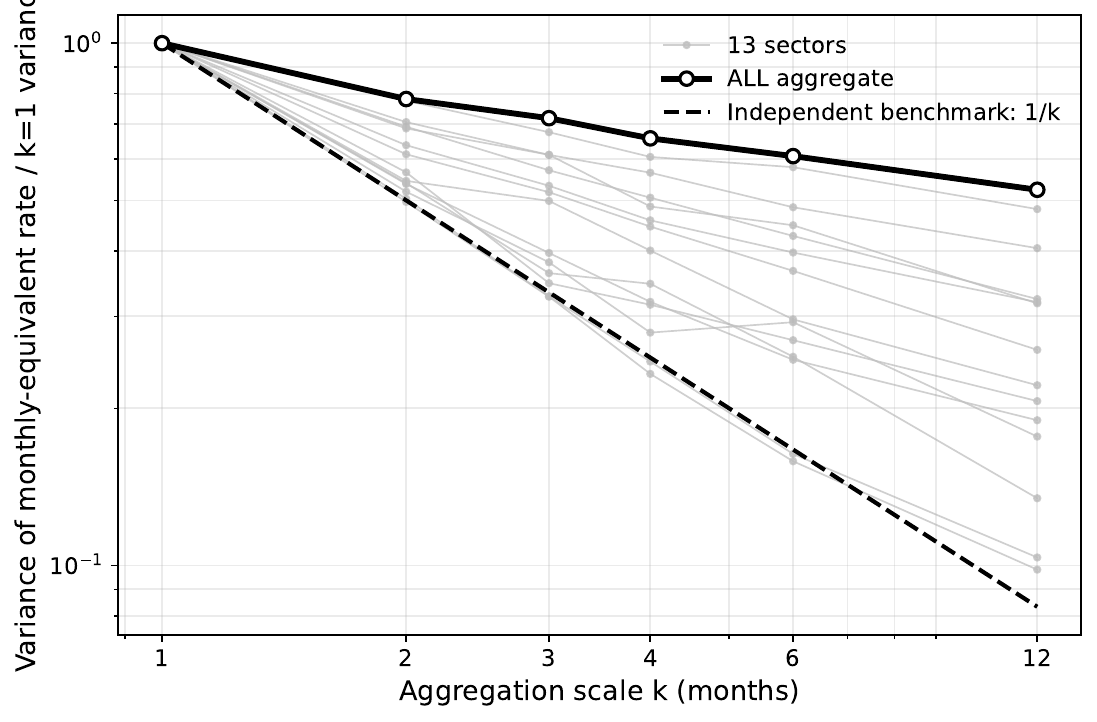}
\caption{Variance scaling of monthly-equivalent default rates.}
\label{fig:variance_scaling}
\end{subfigure}
\hfill
\begin{subfigure}[t]{0.48\textwidth}
\centering
\includegraphics[width=\textwidth]{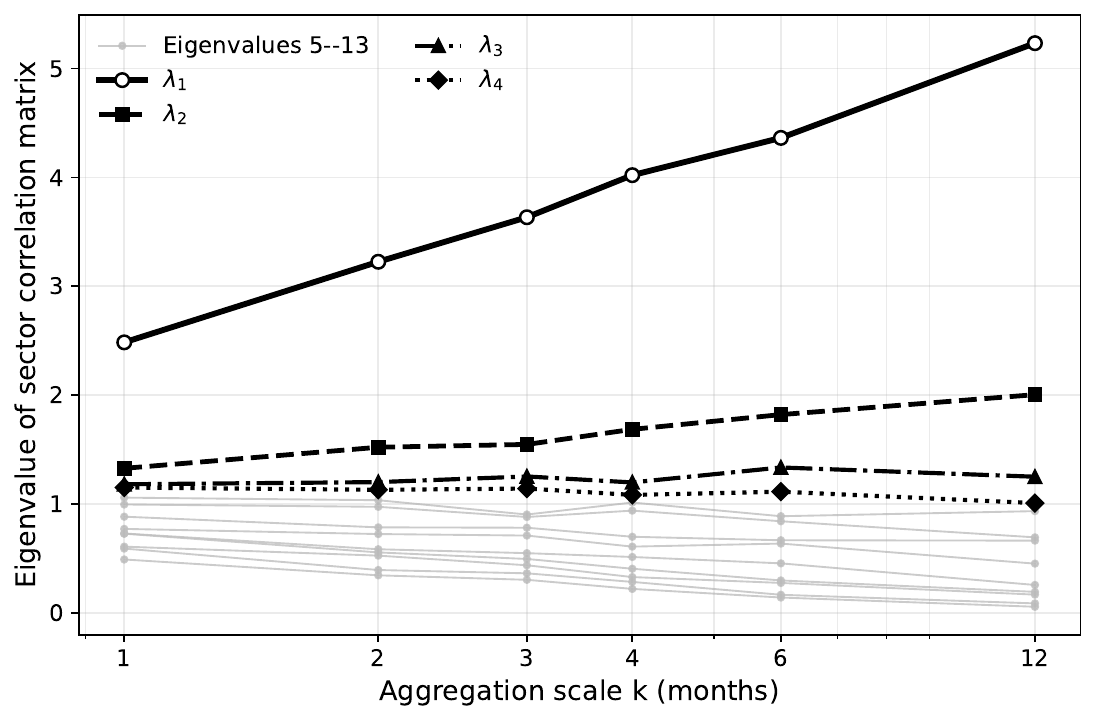}
\caption{Eigenvalue scaling of the sectoral correlation matrix.}
\label{fig:empirical_eigen_scaling}
\end{subfigure}
\caption{
Empirical coarse-graining diagnostics for monthly sector-level default data.
Panel (a) shows variance scaling of monthly-equivalent default rates. Panel (b)
shows eigenvalue scaling of the sectoral correlation matrix. The dashed line in
panel (a) is the independent-month benchmark.
}
\label{fig:empirical_coarse_graining_diagnostics}
\end{figure}

If monthly sectoral default-rate shocks were independent over time, the variance
of the monthly-equivalent rate in Eq.~\eqref{eq:monthly_equiv_rate} would decay
approximately as $1/k$. Empirically, the decay is substantially slower.
Figure~\ref{fig:empirical_coarse_graining_diagnostics}(a) shows the variance
scaling for the 13 sectoral series and for the aggregate ALL series. Thin gray
lines represent individual sectors, while the thick black line represents the
aggregate ALL series. At $k=12$, the variance ratio of the ALL series relative
to the independent benchmark is approximately 2.7. This slow decay indicates
that sectoral default risk is not monthly white noise: high-risk and low-risk
states persist long enough to survive annual aggregation. The particularly slow
decay of the ALL series also reflects the contribution of persistent cross-sector
comovement, because aggregation across sectors preserves common credit-risk
fluctuations while averaging out part of the sector-specific noise.

The empirical correlation matrix also changes with the aggregation horizon. Let
$\widehat C^{(k)}$ denote the sample correlation matrix of the sectoral
monthly-equivalent default-rate vector
$\bm r_b^{(k)}=(r_{b,1}^{(k)},\ldots,r_{b,S}^{(k)})^\top$ over non-overlapping
blocks. Let
\begin{equation}
\widehat C^{(k)}\widehat{\bm v}_j^{(k)}=
\widehat\lambda_j^{(k)}\widehat{\bm v}_j^{(k)},
\qquad
\widehat\lambda_1^{(k)}\ge\widehat\lambda_2^{(k)}\ge\cdots\ge
\widehat\lambda_S^{(k)}
\end{equation}
be its eigenvalue decomposition. 
The leading eigenvalues increase from the monthly to the annual horizon: the
first two empirical eigenvalues are approximately
$\widehat\lambda_1^{(1)}=2.4848$ and $\widehat\lambda_2^{(1)}=1.3276$ at the
monthly scale, while at $k=12$ they are approximately
$\widehat\lambda_1^{(12)}=5.2328$ and $\widehat\lambda_2^{(12)}=2.0043$.
Thus temporal aggregation strengthens the low-rank component of sectoral default
dependence. Figure~\ref{fig:empirical_coarse_graining_diagnostics}(b) reports
the corresponding empirical eigenvalue scaling. 
The fifth and smaller eigenvalues are shown in gray as a background reference,
indicating that the main effect of temporal aggregation is concentrated in the
leading low-rank components.

\subsection{Leading empirical eigenmodes}
\label{subsec:eigenmodes}

\begin{figure}[htbp]
\centering
\includegraphics[width=0.82\textwidth]{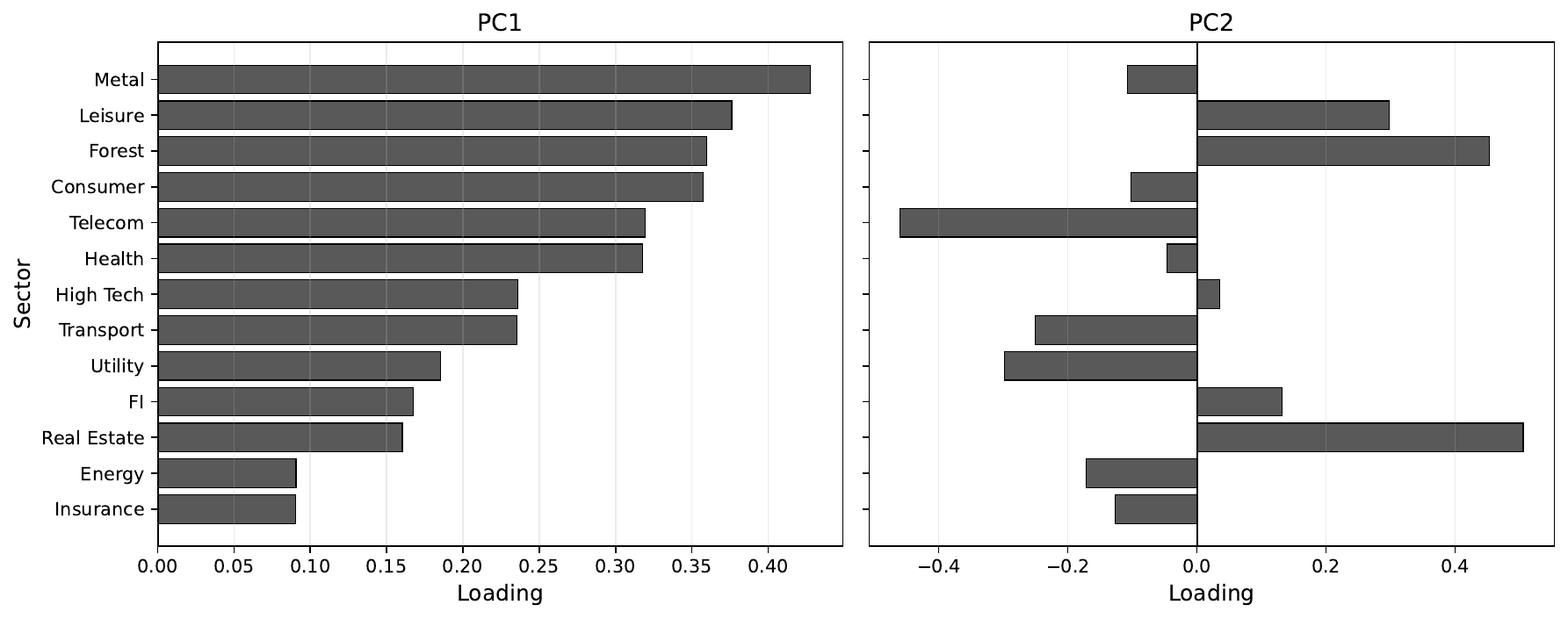}
\caption{Leading empirical eigenvectors of the monthly sectoral correlation
matrix. The first eigenvector has broadly aligned sector loadings and is
interpreted as a market-wide default-risk mode. The second eigenvector changes
the relative allocation of default risk across sectors and is interpreted as a
sector-rotation mode.}
\label{fig:eigenmodes}
\end{figure}

The eigenvalue-scaling diagnostic in Figure~\ref{fig:empirical_eigen_scaling}
shows that temporal aggregation mainly amplifies the leading low-rank components
of sectoral default dependence. We therefore examine the corresponding empirical
eigenvectors at the monthly scale. Figure~\ref{fig:eigenmodes} reports the first
two empirical eigenvectors of $\widehat C^{(1)}$.

The sign of an eigenvector is arbitrary. We orient the first empirical
eigenvector so that its average loading is positive. With this sign convention,
the first empirical eigenvector has broadly aligned sector loadings and is
interpreted as a market-wide default-risk mode. A high value of the corresponding
score then corresponds to higher default probabilities across many sectors at
the same time. This mode is naturally associated with credit-cycle or
macro-financial stress.

The second empirical eigenvector is interpreted as a sector-rotation mode. It
changes the relative allocation of default risk across sectors and can generate
negative or weak dependence between some sector pairs even when the market-wide
factor is positive. This distinction is central for portfolio credit risk: a
one-factor model may be adequate for total default counts, but sector-vector loss
allocation depends on additional directions.

The market-wide and sector-rotation interpretations of these two empirical
directions motivate the two-dimensional latent-factor specification introduced
in the next section.

\subsection{Persistence of empirical principal-component scores}
\label{subsec:pc_score_persistence}

The preceding diagnostics show that the first two empirical eigenvectors are
interpretable directions of sectoral default-rate comovement. We next examine
whether the corresponding empirical score series are persistent over time. This
step is important because it distinguishes the present approach from static PCA:
a PCA decomposition identifies directions in sector space, whereas a dynamic
model requires persistent score dynamics that can be propagated forward and
coarse-grained across horizons.

\begin{figure}[htbp]
\centering
\includegraphics[width=0.62\textwidth]{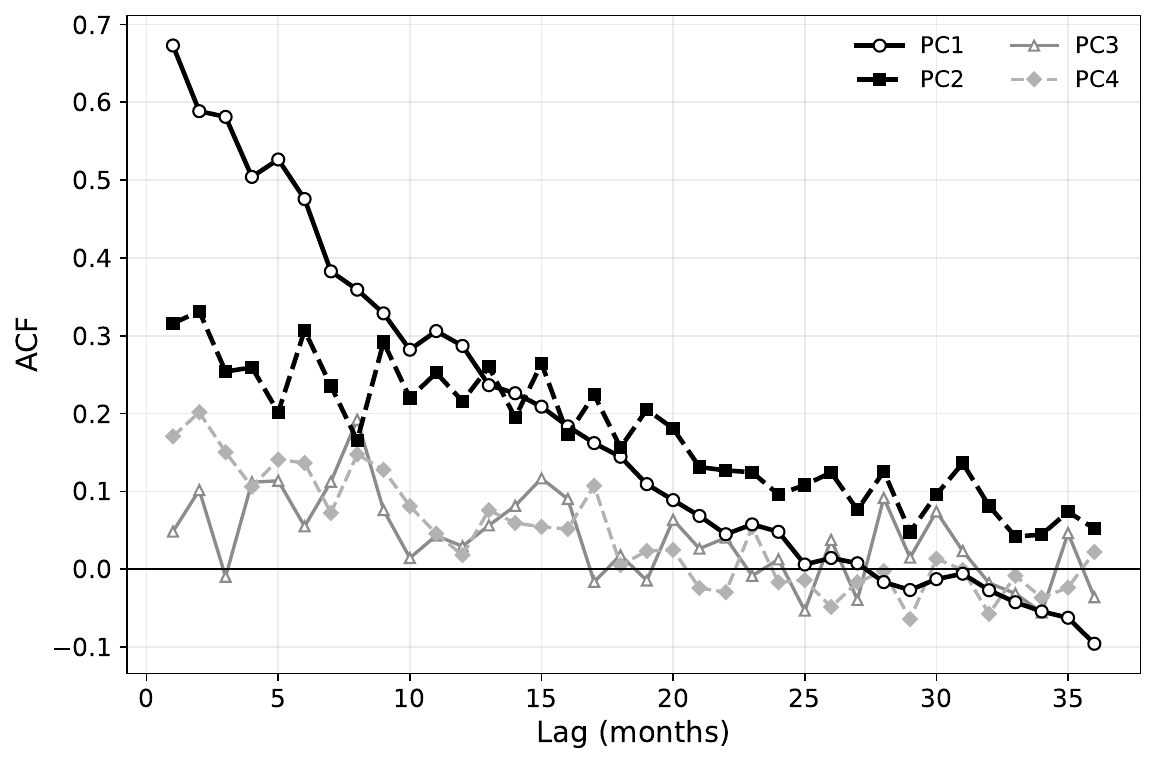}
\caption{
Empirical autocorrelation functions of the first four monthly principal-component
score series at $k=1$.
}
\label{fig:pc_score_acf}
\end{figure}

For each month $t$, define the standardized monthly sectoral default-rate vector
\[
z_{t,s}=\frac{r^{(1)}_{t,s}-\bar r^{(1)}_s}{\hat\sigma_s},
\qquad s=1,\ldots,S,
\]
where $\bar r^{(1)}_s$ and $\hat\sigma_s$ are the sample mean and sample standard
deviation of the monthly default-rate series for sector $s$. We then define the
empirical score associated with the $j$th correlation eigenmode by
\[
\hat f_{j,t}=z_t^\top \hat v^{(1)}_j,
\qquad j=1,\ldots,4.
\]
Here $\hat v^{(1)}_j$ is the $j$th empirical eigenvector of the monthly sectoral
correlation matrix. Thus $\hat f_{j,t}$ is the score obtained by projecting the
standardized monthly sectoral default-rate vector onto the $j$th correlation
eigenmode. Figure~\ref{fig:pc_score_acf} shows the sample autocorrelation functions
of these score series.

The first principal-component score displays strong short-run persistence and
decays gradually over roughly two years. The second principal-component score
has a smaller initial autocorrelation but remains positive over a longer horizon,
with weak persistence visible up to roughly three years. By contrast, the third
and fourth score series fluctuate closer to zero and do not show comparably
stable persistence. These diagnostics indicate that the first two empirical score series contain
persistent low-rank temporal structure, whereas the higher-order score series are
less stable. This provides the empirical motivation for the two-factor dynamic
specification developed below.

\section{Dynamic low-rank AR(1)--Binomial model}
\label{sec:model}

The empirical diagnostics in Section~\ref{sec:data} show that temporal
coarse-graining amplifies the leading low-rank components of sectoral default
dependence and that the first two empirical principal-component score series are
persistent. We therefore use the leading empirical eigenvectors as fixed loading
directions in a dynamic state-space model. The purpose of the model is not
merely to reproduce the monthly correlation matrix, but to infer posterior
sectoral default-probability paths that can be temporally coarse-grained to
longer horizons.

\subsection{Observation equation and low-rank latent state}
\label{subsec:observation_low_rank_state}

Using the notation introduced in Section~\ref{subsec:data_panel}, conditional on
a latent monthly sectoral default probability $p_{t,s}$, the observed default
count is modeled as
\[
L_{t,s}\mid p_{t,s},N_{t,s}\sim \Bin(N_{t,s},p_{t,s}).
\]
The latent default probability is represented on the probit scale:
\[
p_{t,s}=\Phi(y_{t,s}),
\]
where $\Phi$ denotes the standard normal cumulative distribution function.

The probit-scale latent state is decomposed into a sector-specific level, a
low-rank common component, and an idiosyncratic residual:
\[
y_{t,s}=\mu_s+\sum_{r=1}^{R}\lambda_{sr}F_{r,t}+\varepsilon_{t,s}.
\]
Here $\mu_s$ is the sector-specific baseline level, $F_{r,t}$ is the $r$th
latent credit-state factor, $\lambda_{sr}$ is its loading for sector $s$, and
$\varepsilon_{t,s}$ is an idiosyncratic residual.

The loading directions are fixed to the leading empirical eigenvectors of the
monthly sectoral correlation matrix:
\[
\bm\lambda_r=\widehat{\bm v}_r^{(1)},\qquad r=1,\ldots,R,
\]
where $\widehat{\bm v}_r^{(1)}$ is the $r$th empirical eigenvector of
$\widehat C^{(1)}$. These empirical eigenvectors are used only to fix the loading directions.
Their associated empirical eigenvalues are not used to impose amplitude
normalization on the latent factors; the factor amplitudes and persistence are
estimated within the Bayesian state-space model.
Fixing the loading directions removes rotational ambiguity and keeps the
interpretation of the first two factors tied to the market-wide and
sector-rotation modes identified in Section~\ref{subsec:eigenmodes}.

The idiosyncratic residuals are modeled as independent Gaussian shocks with a
common residual scale,
\[
\varepsilon_{t,s}\sim \Normal(0,\sigma_\varepsilon^2),
\]
independent across months and sectors conditional on the latent factors.
The common-residual specification provides a parsimonious way to absorb local
sector-month variation not represented by the persistent low-rank factors,
without introducing weakly identified sector-specific residual scales.

Each common factor follows a stationary AR(1) process,
\[
F_{r,t+1}=\phi_r F_{r,t}+\eta_{r,t+1},
\qquad
\eta_{r,t+1}\sim \Normal(0,\sigma_{\eta,r}^2),
\qquad
0<\phi_r<1,
\]
where the innovation scale $\sigma_{\eta,r}$ is estimated from the data. The
initial state is drawn from the stationary distribution,
\[
F_{r,0}\sim \Normal\left(
0,
\frac{\sigma_{\eta,r}^2}{1-\phi_r^2}
\right).
\]
Under stationarity, the marginal variance of factor $r$ is
\[
\Var(F_{r,t})=\frac{\sigma_{\eta,r}^2}{1-\phi_r^2}.
\]
The persistence parameter $\phi_r$ controls the half-life of sectoral
credit-state shocks,
\[
h_r=\frac{\log(0.5)}{\log(\phi_r)}.
\]

This construction embeds the empirical eigenvectors in a state-space model:
the eigenvectors identify sectoral directions, while the model assigns
persistent stochastic dynamics, observation noise, latent-path uncertainty,
posterior predictive distributions, and a temporal coarse-graining map to those
directions. The factor amplitudes are therefore learned through the posterior
distribution of the innovation scales and persistence parameters, rather than
being fixed by the empirical eigenvalues.

Let
\[
\bm F_t=(F_{1,t},\ldots,F_{R,t})^\top .
\]
The factor process can be written as a diagonal VAR(1) process,
\[
\bm F_{t+1}=A\bm F_t+\bm \eta_{t+1},
\]
where
\[
A=\diag(\phi_1,\ldots,\phi_R),
\qquad
D=\diag(\sigma_{\eta,1},\ldots,\sigma_{\eta,R}),
\qquad
\bm\eta_{t+1}\sim \mathcal{N}(\bm 0,D^2).
\]
Under stationarity, define
\[
V_F=\diag\left(
\frac{\sigma_{\eta,1}^2}{1-\phi_1^2},
\ldots,
\frac{\sigma_{\eta,R}^2}{1-\phi_R^2}
\right).
\]
Then
\[
\Cov(F_{r,t},F_{r',t+h})=
\frac{\sigma_{\eta,r}^2}{1-\phi_r^2}\phi_r^{h}\delta_{r,r'},
\qquad h\ge 0,
\]
where $\delta_{a,b}$ denotes the Kronecker delta.

Let
\[
\bm y_t=(y_{t,1},\ldots,y_{t,S})^\top,
\qquad
\bm\mu=(\mu_1,\ldots,\mu_S)^\top,
\]
and let $\Lambda$ denote the $S\times R$ loading matrix whose $(s,r)$ entry is
$\lambda_{sr}$. The probit-scale latent sectoral state can then be written as
\begin{equation}
\bm y_t=\bm\mu+\Lambda\bm F_t+\bm\varepsilon_t,
\label{eq:latent_state_vector}
\end{equation}
where
\[
\bm\varepsilon_t\sim \Normal(\bm 0,\sigma_\varepsilon^2 I_S).
\]
Conditional on the model parameters, the monthly sectoral latent vector is
Gaussian under stationarity:
\[
\bm y_t\sim
\Normal(\bm\mu,\Omega_0),
\]
where
\[
\Omega_0=\Lambda V_F\Lambda^\top+\sigma_\varepsilon^2 I_S.
\]
More generally, the lag-$h$ cross-covariance matrix of the probit-scale
sectoral state is
\begin{equation}
\Omega_h=\Cov(\bm y_t,\bm y_{t+h})=
\Lambda V_F A^{|h|}\Lambda^\top+
\sigma_\varepsilon^2 I_S \delta_{h,0}.
\label{eq:lag_h_covariance}
\end{equation}
Equivalently, componentwise,
\[
\operatorname{Cov}(y_{t,s},y_{t+h,u})=
\sum_{r=1}^{R}\lambda_{sr}\lambda_{ur}\frac{\sigma_{\eta,r}^2}{1-\phi_r^2}
\phi_r^{|h|}+\sigma_\varepsilon^2 \delta_{h,0}\delta_{s,u}.
\]
For $h=0$, this expression gives the monthly model-implied sectoral covariance
matrix on the probit scale. For $h>0$, the residual term drops out, and
cross-month dependence is generated solely by the persistent low-rank factors.

\subsection{Temporal coarse-graining of latent probability paths}
\label{subsec:temporal_coarse_graining}

The monthly model defined above induces a finite-dimensional Gaussian law over
any block of consecutive months. Temporal coarse-graining is obtained by applying
a survival-aggregation map to this monthly latent path.

For a block of $k$ consecutive months, stack the monthly $S$-dimensional
sectoral state vectors as
\[
\bm y^{(k)}=
\begin{pmatrix}
\bm y_1\\
\vdots\\
\bm y_k
\end{pmatrix}
=(\bm y_1^\top,\ldots,\bm y_k^\top)^\top
\in \mathbb{R}^{kS}.
\]
Conditional on the model parameters,
\[
\bm y^{(k)} \sim \Normal(\bm\mu^{(k)},\Sigma_y^{(k)}),
\]
where
\[
\bm\mu^{(k)}=(\bm\mu^\top,\ldots,\bm\mu^\top)^\top .
\]
The covariance matrix $\Sigma_y^{(k)}$ is a block Toeplitz matrix whose
$(i,j)$ block is
\[
\left[\Sigma_y^{(k)}\right]_{ij}=\Omega_{|i-j|},\qquad i,j=1,\ldots,k,
\]
where $\Omega_h$ is the lag-$h$ cross-covariance matrix defined above. Thus, 
the $k$-month block inherits the low-rank sectoral structure through $\Lambda$, the
factor innovation scales through $V_F$, and the temporal persistence through the
off-diagonal block covariances.

For sector $s$, the $k$-month default probability induced by a monthly
latent path is obtained by survival aggregation,
\[
p_s^{(k)}=
1-\prod_{j=1}^{k}\{1-\Phi(y_{j,s})\}.
\]
We write this componentwise transformation as
\begin{equation}
\bm p^{(k)}=T_k(\bm y^{(k)}),
\qquad
T_k:\R^{kS}\to\R^S.
\label{eq:survival_map_vector}
\end{equation}
The induced $k$-horizon mixing distribution is
\begin{equation}
G_k=\mathcal{L}\left(\bm p^{(k)}\right)=
\mathcal{L}\left(T_k(\bm y^{(k)})\right),
\qquad
\bm y^{(k)}\sim \Normal(\bm\mu^{(k)},\Sigma_y^{(k)}).
\label{eq:induced_mixing_distribution}
\end{equation}
This distribution summarizes how monthly latent credit-state dynamics are
transformed into horizon-$k$ sectoral default probabilities. Conditional on
$\bm p^{(k)}$, sectoral default counts are modeled by independent binomial
observation equations. Hence, dependence among sectoral default counts at horizon
$k$ is generated by the joint mixing distribution $G_k$, not by conditional
dependence in the binomial observation layer.

The same distribution $G_k$ also induces a copula for the coarse-grained sectoral
default-probability vector. In the empirical implementation, $G_k$ and the
corresponding copula are approximated by posterior samples of the
coarse-grained probability vectors. Thus, the posterior-implied copula is a
sample-based approximation to the copula induced by the finite-dimensional
Gaussian AR(1) latent state and the nonlinear survival-aggregation map. This
distinguishes the present construction from a static copula model fitted directly
at horizon $k$: here the horizon-dependent copula is induced by temporally
coarse-graining monthly latent credit-state dynamics.

A finite-dimensional integral representation of the induced mixing distribution,
default-count mixture, and copula is given in
Appendix~\ref{app:coarse_graining_integrals}.

\subsection{Relative amplification of persistent factor components}
\label{subsec:persistent_factor_amplification}

Equation~\eqref{eq:induced_mixing_distribution} defines temporal
coarse-graining through a nonlinear survival-aggregation map. The following
calculation does not replace this nonlinear map; rather, it provides a
probit-scale linear diagnostic for how the underlying AR(1) covariance structure
is amplified by temporal aggregation. For this purpose, we consider the
block-averaged probit-scale latent state
\[
\bar{\bm y}^{(k)}=\frac{1}{k}\sum_{j=1}^k \bm y_j .
\]
Using the lag-$h$ covariance matrix $\Omega_h$ in
Eq.~\eqref{eq:lag_h_covariance}, its covariance matrix is
\begin{equation}
\Omega_{\mathrm{avg}}^{(k)}=\Cov\left(\bar{\bm y}^{(k)}\right)=\frac{1}{k^2}\sum_{i=1}^k
\sum_{j=1}^k\Omega_{|i-j|}.
\label{eq:block_average_covariance}
\end{equation}
Substituting Eq.~\eqref{eq:lag_h_covariance} into
Eq.~\eqref{eq:block_average_covariance} gives
\[
\Omega_{\mathrm{avg}}^{(k)}=\sum_{r=1}^R\frac{\sigma_{\eta,r}^2}{1-\phi_r^2}
g(\phi_r,k)\bm\lambda_r\bm\lambda_r^\top+\frac{\sigma_\varepsilon^2}{k} I_S,
\]
where $\bm\lambda_r=\Lambda\bm e_r$, and
\begin{equation}
g(\phi,k)=\frac{1}{k^2}\sum_{i=1}^k\sum_{j=1}^k\phi^{|i-j|}=
\frac{1}{k}+\frac{2}{k^2}\sum_{h=1}^{k-1}(k-h)\phi^h .
\label{eq:g_phi_k}
\end{equation}
The scalar function $g(\phi,k)$ collects the contribution of temporal
dependence within a $k$-month block. Differentiating
Eq.~\eqref{eq:g_phi_k} with respect to $\phi$,
\[
\frac{\partial g(\phi,k)}{\partial \phi}=\frac{2}{k^2}\sum_{h=1}^{k-1}(k-h)h\phi^{h-1}>0
\]
for $0\leq \phi<1$ and $k>1$. Thus, $g(\phi,k)$ is strictly increasing in
$\phi$. More persistent factors are therefore selectively amplified by
temporal coarse-graining.

Because the loading directions are fixed to orthonormal empirical eigenvectors,
the rank-one factor directions are orthogonal:
\[
\Lambda^\top\Lambda = I_R .
\]
Ignoring the common isotropic residual contribution, the factor-induced
eigenvalue contribution associated with the $r$th loading direction is
\begin{equation}
\mu_r^{(k)}=\frac{\sigma_{\eta,r}^2}{1-\phi_r^2}g(\phi_r,k).
\label{eq:factor_eigenvalue_contribution}
\end{equation}
Hence, for two factors $1$ and $j$,
\[
\frac{\mu_1^{(k)}}{\mu_j^{(k)}}=\frac{\mu_1^{(1)}}{\mu_j^{(1)}}\frac{g(\phi_1,k)}{g(\phi_j,k)} .
\]
Since $g(\phi,k)$ is strictly increasing in $\phi$,
\[
\phi_1>\phi_j\quad\Longrightarrow\quad
\frac{\mu_1^{(k)}}{\mu_j^{(k)}}>\frac{\mu_1^{(1)}}{\mu_j^{(1)}} .
\]
Thus, a more persistent factor becomes relatively more dominant as the
aggregation horizon increases. In particular, if
\[
\phi_1\simeq \phi_2>\phi_j,\qquad j\geq 3,
\]
then the first two persistent components are selectively amplified relative to
lower-rank components.

The large-$k$ asymptotic form makes this mechanism explicit:
\[
g(\phi,k)=\frac{1+\phi}{1-\phi}\frac{1}{k}+O(k^{-2}).
\]
Substituting this into Eq.~\eqref{eq:factor_eigenvalue_contribution} and using
$1-\phi_r^2=(1-\phi_r)(1+\phi_r)$, we obtain
\[
\mu_r^{(k)}=\frac{\sigma_{\eta,r}^2}{(1-\phi_r)^2}\frac{1}{k}+
O(k^{-2}).
\]
The relative asymptotic weight is controlled by $(1-\phi_r)^{-2}$, so highly
persistent factors remain more prominent than less persistent components at
longer aggregation horizons.

The function $g(\phi,k)$ decreases with $k$, as expected for the variance of a
block average. However, for $\phi>0$ it decreases more slowly than the
independent benchmark $1/k$. Equivalently, the relative factor $k g(\phi,k)$
increases with $k$ and converges to $(1+\phi)/(1-\phi)$. Thus, temporal
coarse-graining does not increase the absolute covariance of the block-averaged
latent state. Rather, it increases the relative weight of persistent components
compared with independent or less persistent components. After normalization to
sectoral correlation matrices, this relative persistence appears as spectral
concentration in the leading empirical modes, as observed in
Fig.~\ref{fig:empirical_coarse_graining_diagnostics}(b).
In the empirical application, this mechanism supports the interpretation that
temporal coarse-graining concentrates sectoral dependence into the leading
market-wide and sector-rotation modes.

\subsection{Posterior coarse-graining and predictive counts}
\label{subsec:posterior_coarse_graining}

The monthly state-space model yields posterior samples of monthly sectoral
default-probability paths. For each posterior draw $m$, the $k$-month default
probability for block $b$ and sector $s$ is defined by survival aggregation:
\[
p_{b,s}^{(k,m)}=1-\prod_{j=0}^{k-1}\left(1-p_{bk+j,s}^{(m)}\right).
\]
This is the probability that an obligor present at the beginning of the block
defaults at least once during the $k$-month horizon, conditional on the sampled
monthly probability path.

The long-horizon predictive count equation is
\begin{equation}
\widetilde L_{b,s}^{(k,m)}
\mid
p_{b,s}^{(k,m)},N_{b,s}^{(k)}
\sim
\Bin\left(N_{b,s}^{(k)},p_{b,s}^{(k,m)}\right).
\label{eq:ppc_aggregated_count}
\end{equation}
No independent annual or multi-period latent process is fitted. The
long-horizon probability vector is obtained only by coarse-graining the monthly
posterior path. This restriction links monthly and longer-horizon dependence.

Let
\[
\bm p_b^{(k,m)}=
\left(
p_{b,1}^{(k,m)},\ldots,p_{b,S}^{(k,m)}
\right)^\top
\]
be the coarse-grained sectoral probability vector. The collection
$\{\bm p_b^{(k,m)}\}_{b,m}$ is a posterior sample from a horizon-dependent
mixing distribution over sectoral default-probability vectors. It provides the
simulation-based approximation to the induced distribution $G_k$ defined by the
finite-dimensional Gaussian block law and the survival-aggregation map. In the
following sections, these samples are summarized through effective correlation
matrices, eigenvalue spectra, posterior-implied copulas, and predictive
default-count distributions.

Details of the Bayesian implementation are reported in
Appendix~\ref{app:bayesian_details}. The main text focuses on the model
structure, the temporal coarse-graining map, and the resulting dependence
diagnostics.

\subsection{Posterior-implied rank copula}
\label{subsec:posterior_rank_copula}

The same posterior sample of coarse-grained probability vectors also defines a
rank-based copula at each horizon. For a given horizon $k$, sector $s$, and
posterior block sample $(b,m)$, define
\[
U_{b,s}^{(k,m)}=
\frac{\rank_{s,k}\left(p_{b,s}^{(k,m)}\right)}{M_k+1},
\]
where $\rank_{s,k}(p_{b,s}^{(k,m)})$ denotes the rank of
$p_{b,s}^{(k,m)}$ among all posterior block samples
$\{p_{b',s}^{(k,m')}: b',m'\}$ for the same sector $s$ and horizon $k$, and
$M_k$ is the total number of such samples. The normalization by $M_k+1$ avoids
boundary values equal to zero or one.

The empirical joint distribution of
\[
\bm U_b^{(k,m)}=
\left(
U_{b,1}^{(k,m)},\ldots,U_{b,S}^{(k,m)}
\right)^\top
\]
approximates the posterior-implied rank copula at horizon $k$.

The term posterior-implied copula is used deliberately. The model does not
assume a parametric copula for sector defaults. Instead, rank dependence is
induced by the posterior distribution of the dynamically coarse-grained sectoral
probability vector. Hence the horizon-dependent copula is a consequence of the
monthly low-rank AR(1) dynamics and the nonlinear survival-aggregation map,
rather than an independently fitted static copula at horizon $k$.

\subsection{Factor selection by eigenvalue-scaling diagnostics}
\label{subsec:factor_selection}

We compare fixed-eigenmode specifications with $R=1,2,3,4$ common factors.
The choice of $R$ is based on a coarse-graining criterion: a model fitted at
the monthly scale should reproduce the aggregation-scale behavior of the
leading empirical eigenvalues without attributing excessive persistence to
non-leading modes.

\begin{figure}[htbp]
\centering
\includegraphics[width=0.92\textwidth]{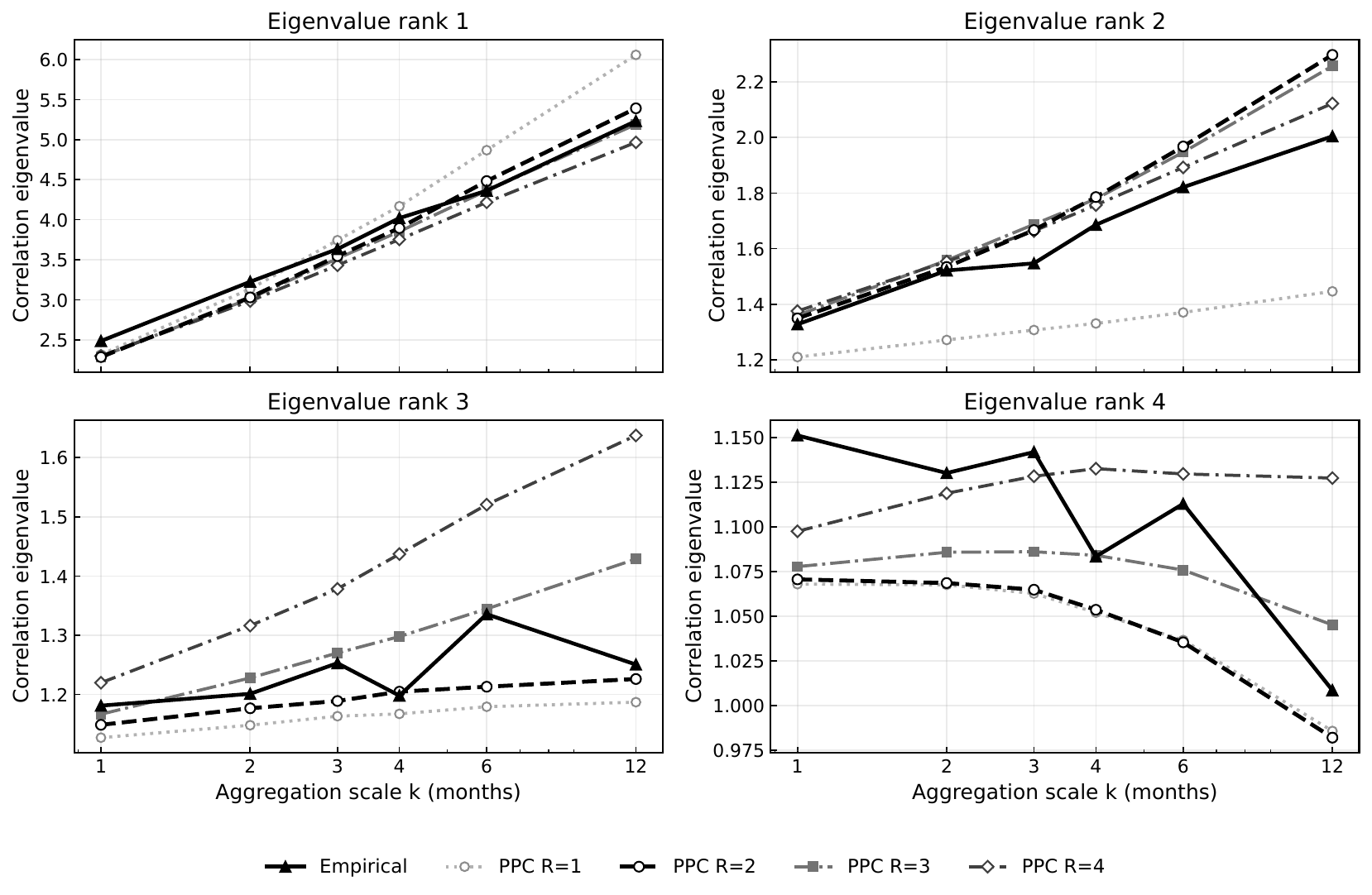}
\caption{
Eigenvalue scaling under temporal coarse-graining for fixed-eigenmode
specifications with $R=1,2,3,4$. The empirical curves are shown by solid
black lines with triangle markers, while posterior predictive medians are
shown by model-specific line styles and markers. The four panels correspond
to eigenvalue ranks 1--4.}
\label{fig:factor_selection_eigen_scaling}
\end{figure}

Figure~\ref{fig:factor_selection_eigen_scaling} compares the empirical and
posterior predictive eigenvalue-scaling curves for the first four eigenvalue
ranks. The $R=1$ specification under-represents the second eigenvalue curve,
whereas higher-rank specifications track selected leading curves more closely.
At the same time, the third and fourth eigenvalue ranks show larger
specification dependence, indicating a trade-off between leading-rank fit and
the stability of non-leading spectral structure.

The empirical eigenvalue scaling in
Figure~\ref{fig:factor_selection_eigen_scaling} provides the target for this
comparison. For each model rank $R$ and aggregation horizon
$k\in\mathcal K=\{1,2,3,4,6,12\}$, let $\widehat{\lambda}^{(k)}_j$ denote the
$j$th empirical eigenvalue and let $\widetilde{\lambda}^{(k)}_{j,R}$ denote the
corresponding posterior predictive median. We compare the normalized
eigenvalue-scaling curves
\[
\widehat{s}^{(k)}_j=
\frac{\widehat{\lambda}^{(k)}_j}{\widehat{\lambda}^{(1)}_j},
\qquad
\widetilde{s}^{(k)}_{j,R}=
\frac{\widetilde{\lambda}^{(k)}_{j,R}}
{\widetilde{\lambda}^{(1)}_{j,R}} .
\]
The eigenvalue-scaling RMSE for rank $j$ is defined as
\[
\mathrm{RMSE}_j(R)=\left[\frac{1}{|\mathcal K|}
\sum_{k\in\mathcal K}\left(\widetilde{s}^{(k)}_{j,R}-\widehat{s}^{(k)}_j\right)^2\right]^{1/2}.
\]
The resulting values are reported in
Table~\ref{tab:factor_selection_diagnostics}.

\begin{table}[htbp]
\centering
\caption{
Factor-selection diagnostics for fixed-eigenmode specifications with
$R=1,2,3,4$ under the common-residual-scale specification.
The table reports eigenvalue-scaling RMSEs for ranks 1--4 over
$k=1,2,3,4,6,12$, together with average eigenvalue-scaling RMSEs over the first
four ranks and over all ranks.
}
\label{tab:factor_selection_diagnostics}
\begin{tabular}{lcccc}
\toprule
Diagnostic & $R=1$ & $R=2$ & $R=3$ & $R=4$ \\
\midrule
Rank 1 eigenvalue-scaling RMSE & 0.2709 & 0.1438 & 0.0987 & 0.0420 \\
Rank 2 eigenvalue-scaling RMSE & 0.1831 & 0.0937 & 0.0742 & 0.0239 \\
Rank 3 eigenvalue-scaling RMSE & 0.0375 & 0.0355 & 0.0812 & 0.1468 \\
Rank 4 eigenvalue-scaling RMSE & 0.0273 & 0.0252 & 0.0498 & 0.0793 \\
\midrule
Average eigenvalue-scaling RMSE, ranks 1--4 & 0.1297 & 0.0746 & 0.0760 & 0.0730 \\
Average eigenvalue-scaling RMSE, all ranks & 0.0790 & 0.0647 & 0.0664 & 0.0641 \\
\bottomrule
\end{tabular}
\end{table}

Moving from $R=1$ to $R=2$ reduces the RMSEs for the first and second
eigenvalue ranks from 0.2709 to 0.1438 and from 0.1831 to 0.0937,
respectively. Increasing the rank further improves some leading-rank
diagnostics; in particular, $R=4$ gives the smallest RMSEs for the first two
ranks. However, the third- and fourth-rank RMSEs are larger for $R=3$ and
$R=4$ than for $R=2$, and the average improvements from higher-rank
specifications are modest.

We next examine the posterior stability of the corresponding AR(1) factor
dynamics. Table~\ref{tab:mcmc_phi_diagnostics} reports posterior summaries and
MCMC diagnostics for the persistence parameters. The first factor is estimated
stably across specifications, with posterior means around 0.95. The second
factor is also persistent in the two-factor specification. By contrast,
additional higher-rank factors are more weakly identified, especially the third
factor in the $R=3$ and $R=4$ specifications, which has lower effective sample
size, elevated $\widehat{R}$, and larger posterior uncertainty.

\begin{table}[htbp]
\centering
\small
\caption{
Posterior summaries and MCMC diagnostics for the AR(1) persistence parameters
in the common-residual-scale fixed-eigenmode models.
}
\label{tab:mcmc_phi_diagnostics}
\begin{tabular}{lrrrrrrr}
\toprule
Model & Factor & Mean & SD & 3\% HDI & 97\% HDI & ESS bulk & $\widehat{R}$ \\
\midrule
$R=1$ & 1 & 0.9555 & 0.0164 & 0.9248 & 0.9858 & 640.5 & 1.008 \\
\midrule
$R=2$ & 1 & 0.9520 & 0.0174 & 0.9192 & 0.9834 & 556.9 & 1.010 \\
& 2 & 0.9527 & 0.0255 & 0.9066 & 0.9928 & 267.3 & 1.009 \\
\midrule
$R=3$ & 1 & 0.9524 & 0.0184 & 0.9182 & 0.9864 & 411.9 & 1.004 \\
& 2 & 0.9484 & 0.0292 & 0.8964 & 0.9934 & 187.3 & 1.025 \\
& 3 & 0.7674 & 0.1546 & 0.5005 & 0.9864 &  71.1 & 1.052 \\
\midrule
$R=4$ & 1 & 0.9496 & 0.0189 & 0.9145 & 0.9851 & 244.5 & 1.021 \\
& 2 & 0.9376 & 0.0345 & 0.8748 & 0.9914 & 223.2 & 1.007 \\
& 3 & 0.6298 & 0.1530 & 0.4034 & 0.9650 & 109.7 & 1.024 \\
& 4 & 0.8297 & 0.0650 & 0.7052 & 0.9367 & 294.5 & 1.021 \\
\bottomrule
\end{tabular}
\begin{flushleft}
\footnotesize
Notes: ESS bulk denotes the bulk effective sample size. The table reports only
the AR(1) persistence parameters. 
\end{flushleft}
\end{table}

Taken together, the eigenvalue-scaling curves, the RMSE diagnostics, and the
posterior stability diagnostics support the use of $R=2$ as the main
specification. The $R=1$ model is too restrictive because it lacks the
sector-rotation component. Higher-rank models improve selected leading-rank
fits, but the gains are modest and are accompanied by less stable higher-rank
factor dynamics and larger errors in some non-leading eigenvalue ranks. The
$R=2$ specification is therefore used as a parsimonious and stable baseline that
captures the market-wide and sector-rotation components without forcing
additional persistent structure into higher-rank modes.

\section{Temporal coarse-graining and posterior predictive diagnostics}
\label{sec:diagnostics}

Having selected the $R=2$ specification in Section~\ref{sec:model}, we now
examine the posterior predictive dependence diagnostics of the main model. The
model is fitted at the monthly scale $k=1$ and is not refitted at the annual
scale. Annual dependence diagnostics are generated by applying the survival
aggregation map to the monthly posterior probability paths. We focus on the
resulting annual correlation matrix, the posterior latent factor trajectory,
posterior-implied rank copulas, and residual sectoral variation.

\subsection{Correlation matrices}

\begin{figure}[htbp]
\centering
\includegraphics[width=0.95\textwidth]{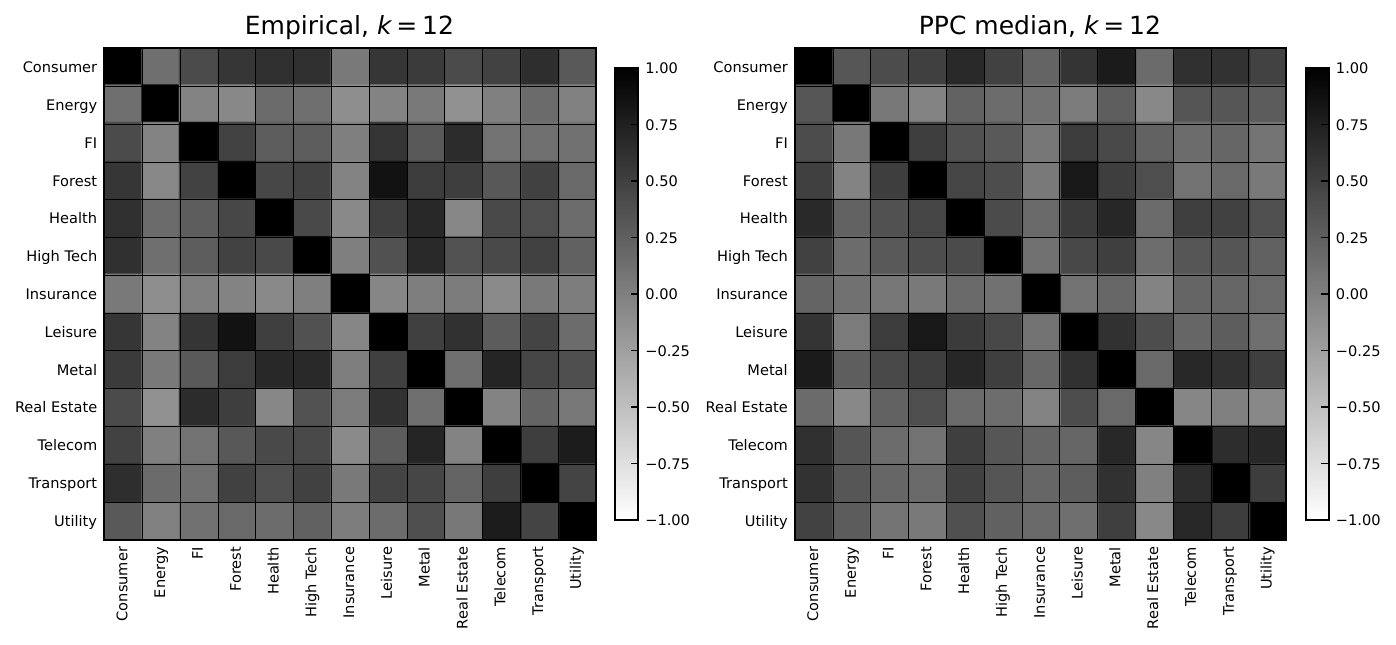}
\caption{
Empirical and posterior predictive sectoral correlation matrices at the annual
horizon $k=12$. The posterior predictive median is obtained from the main
$R=2$ model by survival aggregation of monthly posterior probability paths.
}
\label{fig:corr_k12_main}
\end{figure}

Figure~\ref{fig:corr_k12_main} compares the empirical annual sectoral
correlation matrix with the posterior predictive median under the main $R=2$
specification. 
The posterior predictive annual correlation matrix is generated
by survival aggregation of the monthly posterior probability paths.
The posterior predictive matrix captures the broad block structure of the
empirical annual correlations, although local pairwise discrepancies remain.

\subsection{Latent factor trajectory}

\begin{figure}[t]
\centering
\includegraphics[width=0.78\textwidth]{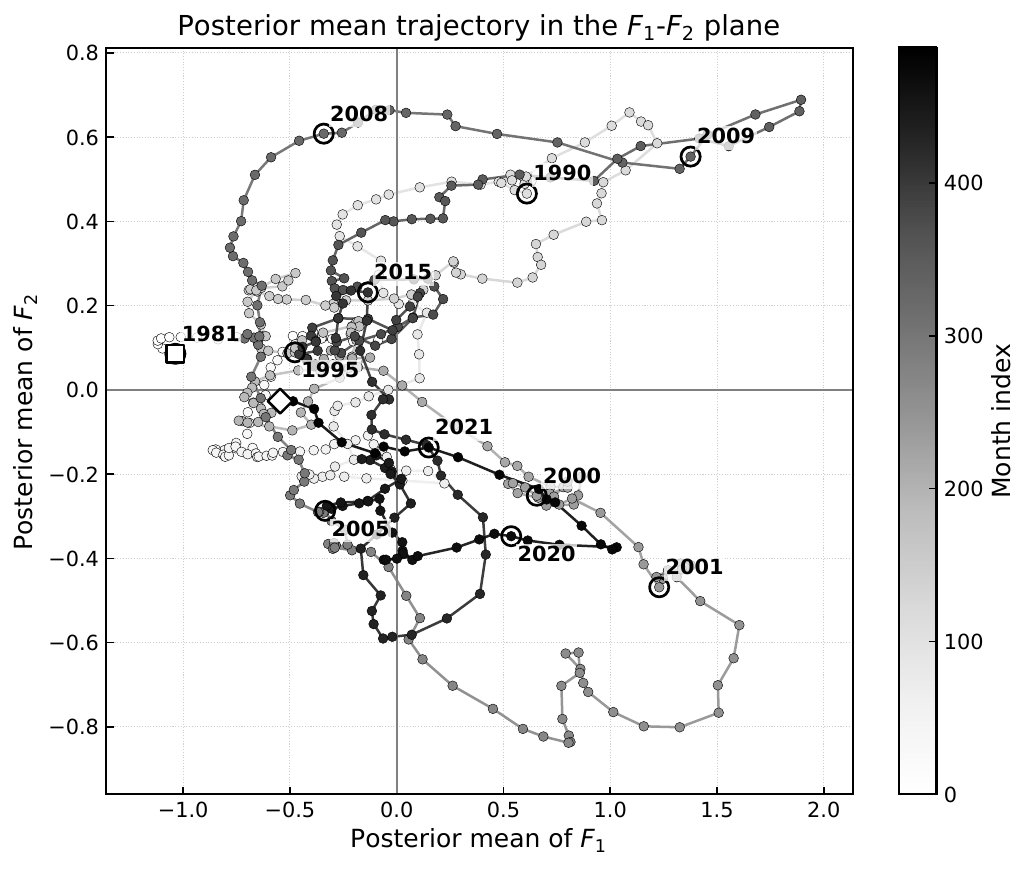}
\caption{
Posterior mean trajectory of the two latent credit-state factors in the
$F_1$--$F_2$ plane. Darker points indicate later months; open circles mark
selected calendar years, and the square marker indicates the beginning of the
sample period.
}
\label{fig:factor_trajectory}
\end{figure}

Figure~\ref{fig:factor_trajectory} shows the posterior mean trajectory of the
two monthly latent credit-state factors in the $F_1$--$F_2$ plane. This is not a
static PCA score plot: the empirical eigenvectors are used only as fixed loading
directions, while the state-space model estimates persistent latent factor paths
through the binomial observation equation and AR(1) dynamics.

The axis interpretation follows from the empirical eigenmodes in
Figure~\ref{fig:eigenmodes}. The first factor represents the market-wide
default-risk mode. The second factor represents sector rotation: positive values
of $F_2$ tilt credit stress toward Real Estate, Forest, Leisure, and FI, whereas
negative values tilt stress toward Telecom, Utility, Transport, Energy, and
Insurance.

This interpretation is consistent with the locations of major stress episodes.
The post-dot-com episode around 2001 lies in a region with high $F_1$ and
negative $F_2$, indicating market-wide stress with a relative tilt toward the
Telecom side. By contrast, the post-Lehman/subprime episode around 2009 lies in
a region with high values of both factors, indicating market-wide stress with a
relative tilt toward the Real Estate--FI side. Thus, the second factor does not
simply measure overall credit stress; it distinguishes the sectoral composition
of stress across episodes.

\subsection{Posterior-implied copulas}

The preceding diagnostics show that the main $R=2$ specification captures the
dominant correlation structure after temporal coarse-graining. We now examine
the induced dependence structure more directly through the posterior-implied
rank copulas defined in Section~\ref{subsec:posterior_rank_copula}. These
copulas are obtained by rank-transforming posterior samples of the
temporally coarse-grained sectoral probability vectors.

For sector pair $(i,j)$, we summarize the posterior-implied rank copula using
the Spearman rank correlation
\[
\rho_{S,ij}^{(k)}=\rho_S\left(U_i^{(k)},U_j^{(k)}\right)
\]
and the upper-tail co-movement ratio at threshold $q$,
\[
\kappa_{ij}^{(k)}(q)=
\frac{\Prob\left(U_i^{(k)}>q,\;U_j^{(k)}>q\right)}{(1-q)^2}.
\]
In the empirical copula diagnostics, we use \(q=0.9\), corresponding to the
upper 10\% marginal rank region.

\begin{figure}[htbp]
\centering
\begin{minipage}{0.78\textwidth}
\centering
\textbf{(a) Metal--Telecom}
\includegraphics[width=\linewidth]{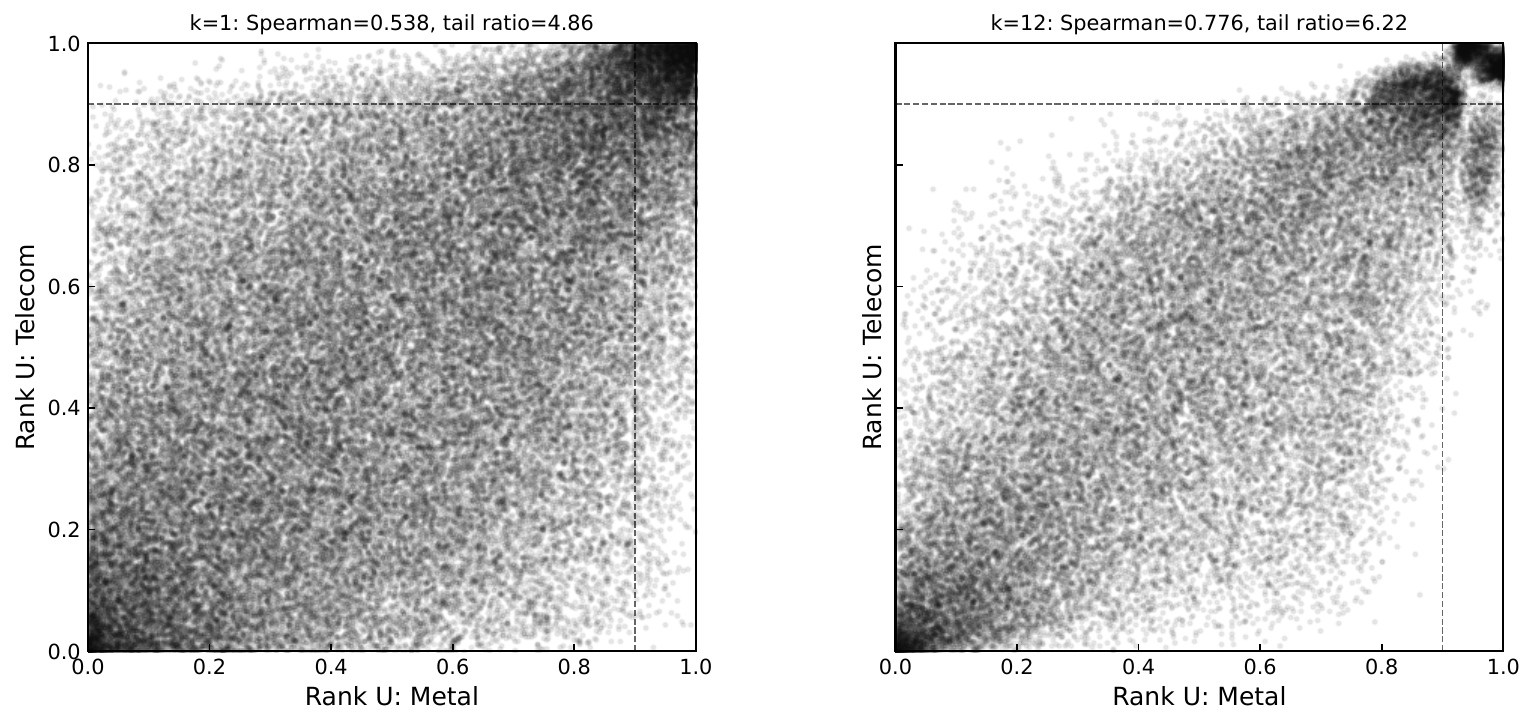}
\end{minipage}

\vspace{0.8em}

\begin{minipage}{0.78\textwidth}
\centering
\textbf{(b) Real Estate--Telecom}
\includegraphics[width=\linewidth]{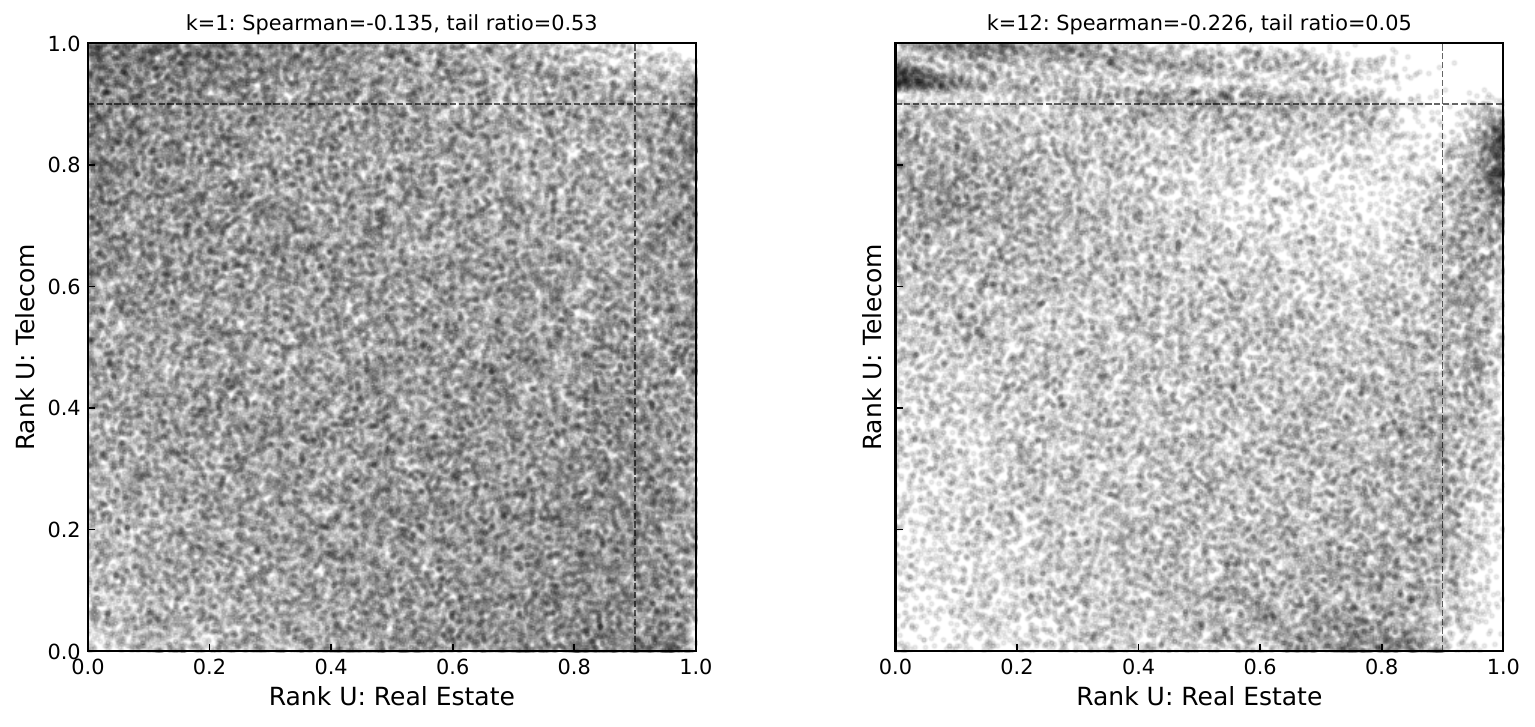}
\end{minipage}
\caption{
Representative posterior-implied rank copulas for selected sector pairs.
Panel (a) shows Metal--Telecom, a pair with positive common-component
covariance. Panel (b) shows Real Estate--Telecom, a pair with negative
common-component covariance. Each pair is shown at the monthly horizon $k=1$
and the annual horizon $k=12$; dashed lines mark the 0.9 marginal rank
threshold.
}
\label{fig:copulas}
\end{figure}

Figure~\ref{fig:copulas} reports two representative sector pairs chosen to
illustrate positive and negative common-component covariance. The Metal--Telecom
pair has positive common-component covariance and displays positive rank
dependence. Temporal coarse-graining strengthens its upper-tail co-movement
from the monthly horizon to the annual horizon.

In contrast, the Real Estate--Telecom pair has negative common-component
covariance and displays tail-exclusion-type dependence. Temporal
coarse-graining suppresses simultaneous upper-tail stress for this pair at the
annual horizon.

These representative examples show that the multi-sector default dependence
induced by the model is not a uniform positive one-factor copula. It is a
horizon-dependent object generated by persistent low-rank latent dynamics and
temporal coarse-graining. The first factor generates broad market-wide stress
co-movement, while the second factor rotates stress across sectors. As a
result, temporal aggregation can sharpen both positive upper-tail co-movement
and tail-exclusion-type dependence, depending on the sector pair.

\subsection{Residual sectoral variation}

The preceding diagnostics show that the $R=2$ fixed-eigenmode model captures the
dominant low-rank dependence structure after temporal coarse-graining. However,
it does not reproduce all sector-specific marginal variance components. The
sector-wise variance decomposition is reported in
Table~\ref{tab:variance_decomposition} in Appendix~\ref{app:variance_decomposition}.

The posterior predictive observed variance is close to the empirical variance
for many sectors, including Energy, Forest, Metal, Telecom, and Transport, and
is moderately above the empirical variance for Consumer and Leisure. The
strongest under-representation occurs in Insurance and Real Estate, where the
posterior-predictive-to-empirical variance ratios are approximately $0.15$ and
$0.36$, respectively. Utility also shows moderate under-representation, with a
ratio of approximately $0.67$. High Tech is not severely under-represented in
posterior predictive observed variance, with a ratio of approximately $0.81$,
but its persistent low-rank latent component accounts for only a small fraction
of the empirical variance.

The decomposition also clarifies the role of the binomial observation layer.
For several sectors, including Energy, Health, High Tech, Transport, and Real
Estate, binomial noise accounts for a large fraction of the model-implied
latent-plus-observation variance. By contrast, Consumer, Forest, Leisure, Metal,
and Telecom receive a more substantial contribution from the persistent
low-rank latent component. Because the common-residual specification uses 
a single residual scale $\sigma_\varepsilon$ for all sectors,
this heterogeneity reflects the interaction between sector exposures, empirical
loading directions, latent factor dynamics, and binomial sampling noise, rather
than sector-specific residual-scale parameters.

We do not interpret the remaining discrepancies as evidence that additional
stable common factors are necessarily required. In the factor-selection
diagnostics, increasing the rank to \(R=3\) or \(R=4\) does not systematically
remove these sector-level variance discrepancies, while it risks over-amplifying
non-leading eigenvalue components. We also found that sector-specific residual
scales did not provide a stable resolution, because the corresponding fits were
less stable and did not robustly eliminate the under-representation. The
discrepancies in Insurance, Real Estate, and High Tech are therefore best viewed
as limitations of the parsimonious two-factor common-residual specification,
suggesting that their resolution would require a more explicit model of
sector-specific dynamics rather than a simple higher-rank or residual-scale
extension.

\section{Forecast evaluation}
\label{sec:forecast}

We now evaluate whether the temporally coarse-grained dynamic factor structure
improves annual default-count forecasts. The comparison focuses on annual
default-count forecasts at horizon $k=12$, using both portfolio-level and
sector-vector predictive distributions. We compare static binomial and
beta-binomial baselines with one-factor and two-factor dynamic specifications.

\subsection{Forecast design and competing models}
\label{subsec:forecast_design}

The preceding section was an in-sample aggregation diagnostic: the monthly model was fitted to the full sample, and posterior monthly probability paths were aggregated to examine annual correlation matrices, eigenvalue scaling, and posterior-implied copulas. We now ask whether the same dynamic low-rank structure also has predictive value. For this purpose, we conduct a rolling 12-month-ahead forecast evaluation of annual default counts, using both portfolio-level and sector-vector predictive distributions.

Let $W$ denote the training-window length and let $H=12$ denote the forecast horizon. At each forecast origin, a model is fitted or calibrated using only the previous $W$ months of observations. The fitted model then generates predictive paths of monthly sectoral default probabilities,
\[
p_{t+1,s},p_{t+2,s},\ldots,p_{t+H,s},
\]
which are converted into an $H$-month default probability by survival aggregation,
\[
p_{t:t+H,s}=1-\prod_{h=1}^{H}{1-p_{t+h,s}}.
\]
Conditional on these probabilities and beginning-of-horizon exposures, sectoral default counts are generated from the binomial observation layer. Forecast performance is evaluated for both the total portfolio default count and the full sector-vector default count.

We compare four forecasting specifications, labelled B0--B3. B0 is a sector-specific constant-rate binomial model. Because its performance depends strongly on the rolling-window length, we select the best window ex post from
\[
W\in\{12,24,36,60,120,180,240\}.
\]
This uses information from the evaluation sample and is therefore an oracle choice rather than a genuine real-time forecasting rule. We deliberately give this advantage to the static binomial baseline, so B0 should be interpreted as a favored static benchmark. B1 is a sector-specific constant-rate beta-binomial model with the same oracle window-selection convention. It allows static overdispersion and is therefore a stronger static baseline than B0.

B2 and B3 are dynamic Bayesian AR(1)--Binomial models with fixed empirical eigenmode loadings. B2 is the one-factor dynamic model, whereas B3 is the two-factor dynamic model. In both cases, monthly latent factors are propagated forward according to their estimated AR(1) dynamics, and survival aggregation maps the predicted monthly probabilities to 12-month predictive count distributions. The rolling forecast implementation is summarized separately in Appendix~\ref{app:forecast_bayesian_setup}.

To avoid look-ahead bias, the fixed eigenmode loading directions used in B2 and B3 are re-estimated separately at each forecast origin using only the corresponding training window. In particular, the sectoral correlation matrix, its leading eigenvectors, and the associated fixed loading directions are computed from the previous $W$ months only. No observations from the forecast horizon or from later months enter the construction of the forecast loading directions.

Forecast quality is evaluated by log predictive score, continuous ranked probability score (CRPS), 90\% predictive interval coverage, predictive standard deviation, and sector-vector log predictive score. The CRPS is a proper scoring rule for distributional forecasts \citep{GneitingRaftery2007,GneitingKatzfuss2014}. 
For a predictive distribution $F$ and observation $y$, it is defined as
\[
\mathrm{CRPS}(F,y)=\mathbb{E}_{F}|X-y|-\frac{1}{2}\mathbb{E}_{F}|X-X'|,
\]
where $X$ and $X'$ are independent draws from $F$. Smaller CRPS is better, whereas larger log predictive scores and sector-vector log predictive scores are better. 
Details of the Bayesian prior specification, posterior sampling settings,
posterior predictive simulation, and Monte Carlo computation of forecast scores
are reported in Appendix~\ref{app:forecast_bayesian_setup}.

\subsection{Portfolio-level forecast comparison}

\begin{figure}[htbp]
\centering
\includegraphics[width=\textwidth]{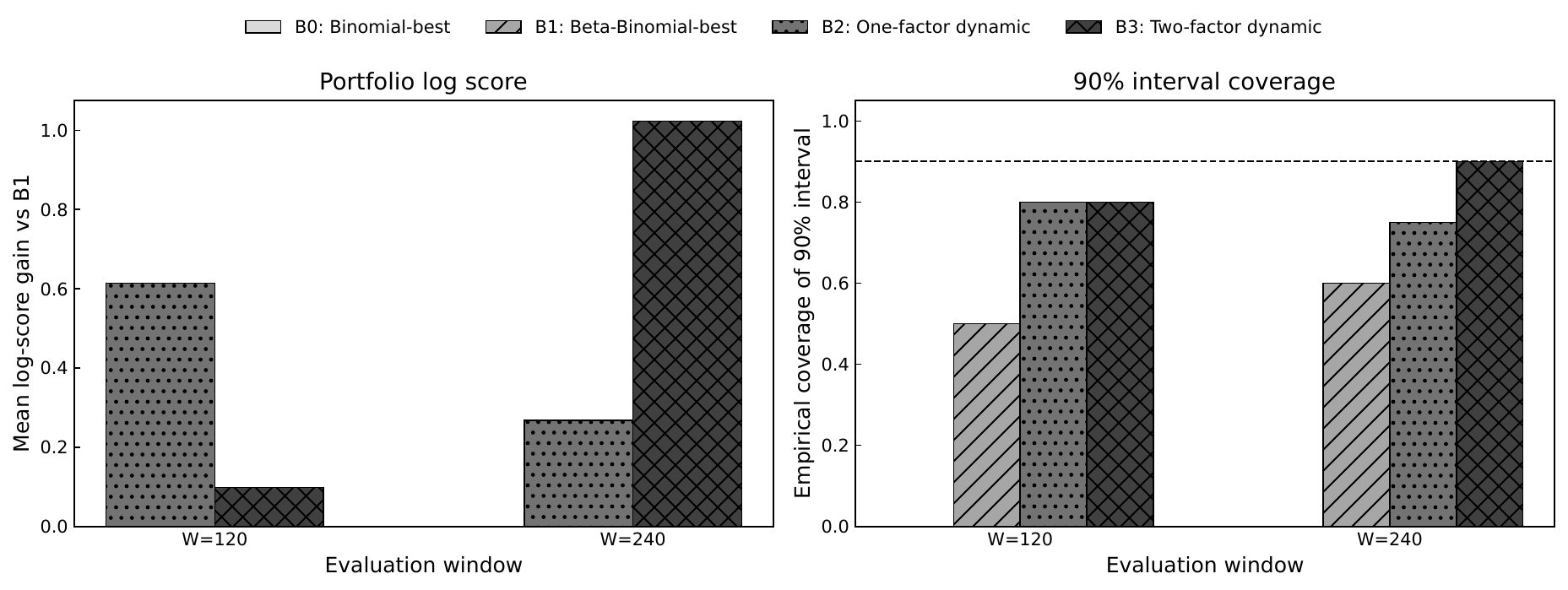}
\caption{
Annual portfolio-level out-of-sample forecast comparison. The left panel
reports portfolio log-score gains relative to the static beta-binomial baseline
B1; B1 is therefore the zero baseline, and B0 is excluded from the left panel
because its log-score gains are far outside the plotted range. The right panel
reports empirical coverage of nominal 90\% predictive intervals; the dashed
line marks the nominal 90\% coverage level.
}
\label{fig:portfolio_forecast}
\end{figure}

Figure~\ref{fig:portfolio_forecast} summarizes the portfolio-level annual
out-of-sample forecast comparison. The left panel reports portfolio log-score
gains relative to the static beta-binomial baseline B1, and the right panel
reports empirical coverage of nominal 90\% predictive intervals. The full
numerical results, including CRPS, predictive standard deviations, and
sector-vector log scores, are reported in Appendix~\ref{app:forecast_tables}.

The static binomial baseline B0 performs poorly even though it is given the
advantage of ex-post oracle window selection. Its log score is far below that of
B1, and its 90\% predictive interval coverage is zero in both evaluation
windows. This indicates that the binomial baseline produces annual predictive
distributions that are much too narrow. The beta-binomial baseline B1 improves
this behavior by introducing static overdispersion, but its empirical coverage
still remains substantially below the nominal 90\% level.

The dynamic models B2 and B3 provide better calibrated annual predictive
distributions. Their 90\% interval coverage is closer to the nominal level than
that of B1, and their predictive standard deviations are substantially larger
than those of the static binomial baseline. Thus, the improvement is not merely
a consequence of static overdispersion; it also reflects predictive dispersion
generated by persistent monthly latent credit-state dynamics propagated over a
12-month forecast horizon.

At the portfolio-total level, the dynamic models substantially improve over the
static baselines. The one-factor dynamic model B2 gives the largest
portfolio-log-score gain for $W=120$ and the smallest CRPS in both evaluation
windows. The two-factor dynamic model B3 gives the largest portfolio-log-score
gain for $W=240$ and attains nominal 90\% coverage in that window. Thus, the
portfolio-level evidence indicates that persistent monthly credit-state dynamics
improve annual aggregate default-count forecasts, although the relative ranking
of B2 and B3 depends on the scoring criterion and the evaluation window.

\subsection{Sector-vector forecast comparison}

We next examine forecast performance for the full sectoral default-count vector.
This comparison is important because the portfolio-level total count can hide
sectoral allocation errors. A model may forecast the aggregate number of
defaults reasonably well while assigning default risk to the wrong sectors. The
sector-vector evaluation therefore provides a more direct test of how the
dynamic factor structure affects the allocation of credit stress across sectors.

\begin{figure}[htbp]
\centering
\includegraphics[width=0.82\textwidth]{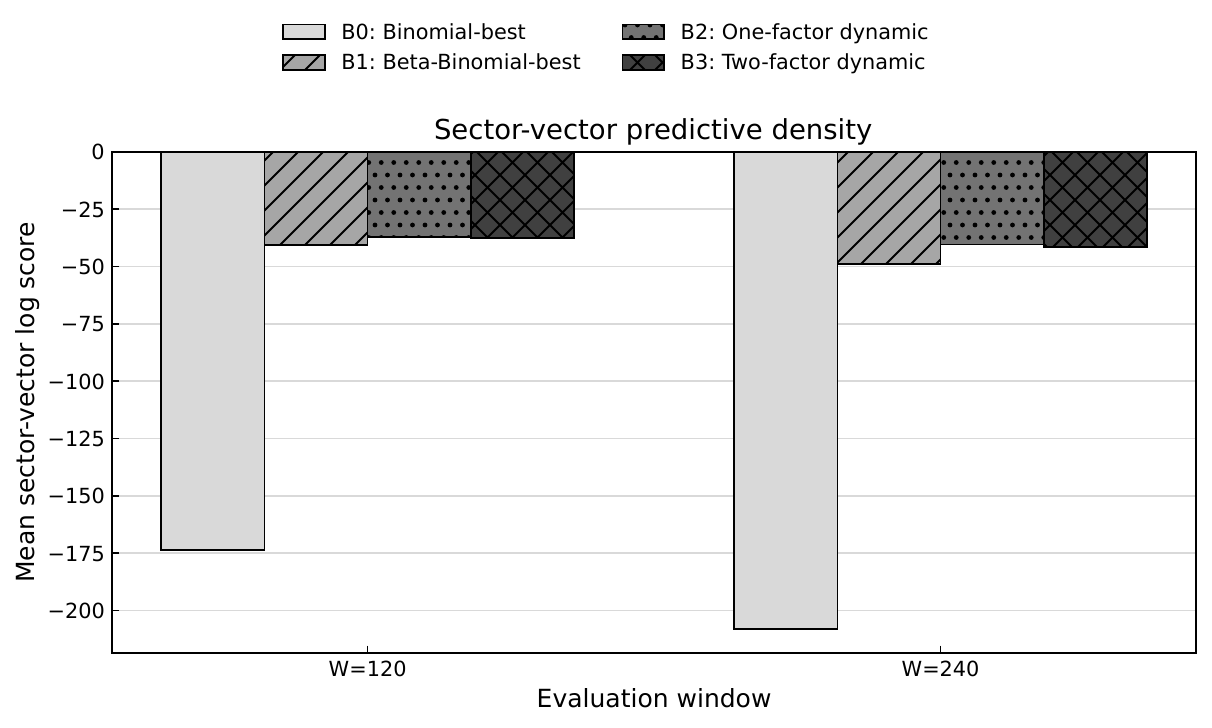}
\caption{
Sector-vector forecast comparison. Bars report mean sector-vector log scores
for the four forecasting models. Larger values are better.
}
\label{fig:sector_vector_forecast}
\end{figure}

Figure~\ref{fig:sector_vector_forecast} compares the sector-vector predictive
density across the four forecasting models. The figure reports mean
sector-vector log scores, for which larger values indicate better predictive
performance. The full numerical comparison, including mean per-sector log scores
and mean per-sector CRPS, is reported in
Table~\ref{tab:sector_vector_forecast}.

The static binomial baseline B0 performs poorly, reflecting narrow predictive
distributions and the absence of dynamic cross-sector dependence. The
beta-binomial baseline B1 substantially improves the sector-vector log score by
allowing static overdispersion, but it remains a sector-by-sector static
benchmark and does not model dynamic low-rank movements in sectoral default
risk.

The dynamic models improve the sector-vector forecast relative to the static
baselines. In log-score-based comparisons, the one-factor dynamic model B2 gives
the best sector-vector log score and the best mean per-sector log score in both
evaluation windows. The two-factor dynamic model B3 is close to B2 in these
log-score measures, but it does not uniformly dominate B2. By contrast, B3 gives
the smallest mean per-sector CRPS in both evaluation windows, indicating an
improvement in this aspect of sector-level distributional calibration.

These results refine the interpretation of the second factor. The first factor
captures the dominant market-wide component of annual default risk and is
already highly effective for both portfolio-total forecasts and log-score-based
sector-vector predictive density. The second factor does not uniformly improve
sector-vector log scores in the rolling forecast experiment. Its contribution
is more limited: it improves per-sector CRPS and affects the calibration of
sectoral default allocations. This is consistent with the interpretation of the
second factor as a sector-rotation component, while also showing that its
forecast benefit depends on the scoring rule.

\section{Discussion}
\label{sec:discussion}

The proposed model is deliberately built on empirical eigenvectors, but their
role is limited to fixing the loading directions. The probabilistic structure is
supplied by the binomial observation layer, common-scale idiosyncratic residual,
AR(1) latent dynamics, posterior state uncertainty, and survival aggregation map.
Thus, the model should be viewed not as a static PCA approximation, but as a
Bayesian state-space model that turns empirical correlation directions into
horizon-dependent predictive distributions for sectoral default probabilities.

\subsection{Relation to dynamic factor models}

Dynamic factor models are a natural tool for default-count and credit-risk data.
They can represent common macro-financial variation, latent frailty, and
sector-specific or industry-level effects, and they are often used for
conditional default-probability estimation and forecasting. The purpose of the
present paper is therefore not to introduce a new general dynamic factor model
for corporate defaults.

The contribution is more specific. We use a deliberately parsimonious
fixed-eigenmode state-space model to study how low-rank credit-state dynamics
estimated at the monthly scale are transformed into annual multi-sector
dependence structures. The fixed loading directions remove rotational ambiguity,
while the common-residual-scale specification avoids over-interpreting weakly
identified sector-specific residual variances.

From this perspective, the model is best interpreted as a scale-consistent
baseline rather than as a fully general default-intensity model. Its value lies
in showing what annual-scale dependence, posterior-implied copula structure, and
forecast dispersion can already be generated by persistent monthly low-rank
dynamics combined with a binomial observation layer, a common-scale
idiosyncratic residual, and survival aggregation. The forecast exercises further 
indicate that this scale-consistent construction
has predictive content, even without introducing a fully specified
sector-specific intensity model.

\subsection{Temporal coarse-graining and effective dependence}

Static copula and factor models usually specify dependence directly at the
horizon of interest. If annual defaults are modeled, an annual copula or annual
correlation matrix is fitted. The present approach takes the opposite route. It
fits the latent dynamics at the monthly scale and derives the annual dependence
structure by temporal coarse-graining. The annual correlation matrix, eigenvalue
spectrum, and posterior-implied rank copula are therefore effective quantities
generated by the monthly latent probability path.

This distinction matters for model interpretation. A high annual correlation
does not necessarily imply strong instantaneous default dependence within a
year. It can arise because the monthly sectoral default-probability vector is
persistent, so that high-risk monthly states cluster inside annual blocks. 
The same mechanism was visible in the aggregate default-count analysis of
\citet{Mori2026TemporalCoarseGraining}; the present paper shows that it also
operates in the sectoral dependence matrix, eigenvalue spectrum, and
posterior-implied rank copula.

Thus, annual dependence parameters should not automatically be interpreted as
primitive one-period dependence parameters. They may instead summarize the
combined effect of monthly persistence, sectoral factor structure, residual
variation, binomial sampling noise, and temporal aggregation, consistent with
the effective-dependence interpretation developed in
\citet{Mori2026ContagionMacro}.

For portfolio risk management, tail-risk measures such as VaR and expected shortfall are functionals of a predictive loss distribution. They cannot be evaluated in a dynamically consistent forward-looking way unless the joint distribution of future sectoral default probabilities and default counts is itself specified. In a multi-sector credit portfolio, this means that forecasting marginal default probabilities is not sufficient: one must also forecast the dependence structure that determines which sectors are likely to enter high-default states together.

The posterior-implied copula in the present framework should therefore be interpreted as a dynamic forward object. Given the current posterior distribution of the monthly latent credit-state factors, the model propagates sectoral default probabilities forward, applies survival aggregation to the target horizon, and generates a predictive distribution of sectoral default-probability vectors and default counts. Tail-risk measures can then be computed from this predictive distribution. The key point is that the copula used for tail-risk calculation is not fitted independently at the annual horizon; it is predicted from monthly latent dynamics. Thus, the model provides a scale-consistent dynamic copula generator for portfolio loss simulation and stress testing,
rather than a static copula calibrated separately at each horizon.

\subsection{Limitations and possible extensions}

The $R=2$ model leaves residual variance in some sectors, most notably
Insurance and Real Estate, and the persistent low-rank latent contribution is
also weak for High Tech. For Insurance and Real Estate, the persistent low-rank
latent component accounts for only a small fraction of the empirical marginal
variance. We treat this remaining variation as local sector-specific variation
rather than forcing it into higher-rank persistent factors. This choice keeps the
model stable and interpretable, but it also limits local sector-level fit.

Several extensions are possible, but the results suggest that they should be
introduced with explicit structure rather than as unrestricted increases in
flexibility. Additional observation-layer overdispersion, such as a beta-binomial
layer, may help describe sectors with excess variance. Sector-specific residual
dynamics could capture persistent local risk not explained by the two common
factors. More flexible specifications could also allow the loading directions or
factor volatilities to vary over long historical periods. We do not include these
extensions in the main specification because the goal of the paper is to isolate
the effect of stable low-rank monthly dynamics and temporal coarse-graining.

\section{Conclusion}
\label{sec:conclusion}

This paper studied how persistent monthly credit-state dynamics generate
annual-scale multi-sector default dependence. We proposed a parsimonious
fixed-eigenmode AR(1)--Binomial state-space model for monthly sectoral default
counts. The empirical eigenvectors identify interpretable sectoral directions:
a market-wide default-risk mode and a sector-rotation mode. Monthly latent
probability paths were then mapped to annual horizons by survival aggregation.

The main finding is that annual dependence should be interpreted as an
effective quantity generated by monthly dynamics and temporal aggregation. The
two-factor model captures the main annual correlation and eigenvalue-scaling
diagnostics without fitting a separate annual dependence model. The
posterior-implied rank copulas further show that temporal aggregation can
sharpen both positive tail co-movement and tail-exclusion patterns, depending
on the sector pair.

The out-of-sample forecast comparison supports the same interpretation. Dynamic
models generate better calibrated annual predictive distributions than favored
static baselines. At the portfolio level, the first factor explains much of the
variation in total annual default counts, so the one-factor dynamic model is
highly competitive for aggregate forecasts. The two-factor model improves some
aspects of forecast calibration, including per-sector CRPS, but it does not
uniformly dominate the one-factor model in log-score-based forecast comparisons.
Thus, the second factor is best interpreted as a sector-rotation component whose
forecast contribution depends on the evaluation criterion.

Overall, the results show that a stable low-rank monthly model, combined with a
binomial observation layer, a common-scale idiosyncratic residual, and survival
aggregation, can generate rich horizon-dependent default dependence. This
provides a scale-consistent baseline for interpreting multi-sector default
correlations, posterior-implied rank copulas, and annual default-count
forecasts.

\clearpage

\appendix

\section{Additional empirical diagnostics}
\label{app:additional_empirical_diagnostics}

This appendix reports additional descriptive diagnostics for the monthly
multi-sector default-count panel used in Section~\ref{sec:data}. The main text
focuses on the data construction and the empirical coarse-graining diagnostics.
Here we provide sector-level summary statistics and a quarterly heatmap that
visualizes the timing and sectoral heterogeneity of default-stress episodes.

\begin{table}[htbp]
\centering
\small
\caption{Sector-level summary statistics for the monthly default-count panel.}
\label{tab:sector_summary_statistics}
\begin{tabular}{lrrrrrrr}
\toprule
Sector & $\bar N$ & $N_{\min}$ & $N_{\max}$ & $L_{\mathrm{tot}}$
& $\bar r_{\mathrm{m}}$ & $\mathrm{sd}(r_{\mathrm{m}})$ & $\hat r_{\mathrm{ann}}$ \\
\midrule
Consumer    &  441.7 &  226 &  757 &  443 & 0.001960 & 0.002672 & 0.023522 \\
Energy      &  249.1 &   89 &  532 &  374 & 0.002406 & 0.004530 & 0.028877 \\
FI          &  677.0 &   87 & 1117 &  179 & 0.000528 & 0.001313 & 0.006334 \\
Forest      &  142.2 &   65 &  223 &  140 & 0.002009 & 0.005024 & 0.024106 \\
Health      &  244.6 &   69 &  483 &  138 & 0.001078 & 0.002534 & 0.012937 \\
High Tech   &  157.9 &   61 &  335 &   78 & 0.001096 & 0.003290 & 0.013146 \\
Insurance   &  494.0 &   33 &  805 &   59 & 0.000370 & 0.002322 & 0.004443 \\
Leisure     &  242.3 &   43 &  387 &  326 & 0.002613 & 0.004677 & 0.031351 \\
Metal       &  392.1 &  217 &  589 &  321 & 0.001613 & 0.002984 & 0.019358 \\
Real Estate &  135.3 &   10 &  310 &   29 & 0.000858 & 0.005810 & 0.010297 \\
Telecom     &  149.4 &   67 &  236 &  161 & 0.001770 & 0.004502 & 0.021246 \\
Transport   &  152.4 &   90 &  237 &  129 & 0.001727 & 0.003645 & 0.020721 \\
Utility     &  434.2 &  235 &  609 &   74 & 0.000316 & 0.001205 & 0.003791 \\
\midrule
ALL         & 3912.2 & 1342 & 6287 & 2451 & 0.001231 & 0.001161 & 0.014768 \\
\bottomrule
\end{tabular}
\begin{flushleft}
\footnotesize
Notes: $\bar N$, $N_{\min}$, and $N_{\max}$ denote the mean, minimum, and maximum
monthly numbers of obligors. $L_{\mathrm{tot}}$ is the total number of defaults.
$\bar r_{\mathrm{m}}$ and $\mathrm{sd}(r_{\mathrm{m}})$ denote the mean and standard
deviation of the monthly default rate. $\hat r_{\mathrm{ann}}$ is the annualized
default rate, computed as $12\bar r_{\mathrm{m}}$. The ALL row is an aggregate
series constructed by summing defaults and exposures across the 13 sectors. The
main multi-sector analysis uses the 13 sectoral series.
\end{flushleft}
\end{table}

\begin{landscape}
\begin{figure}[p]
\centering
\includegraphics[width=0.95\linewidth]{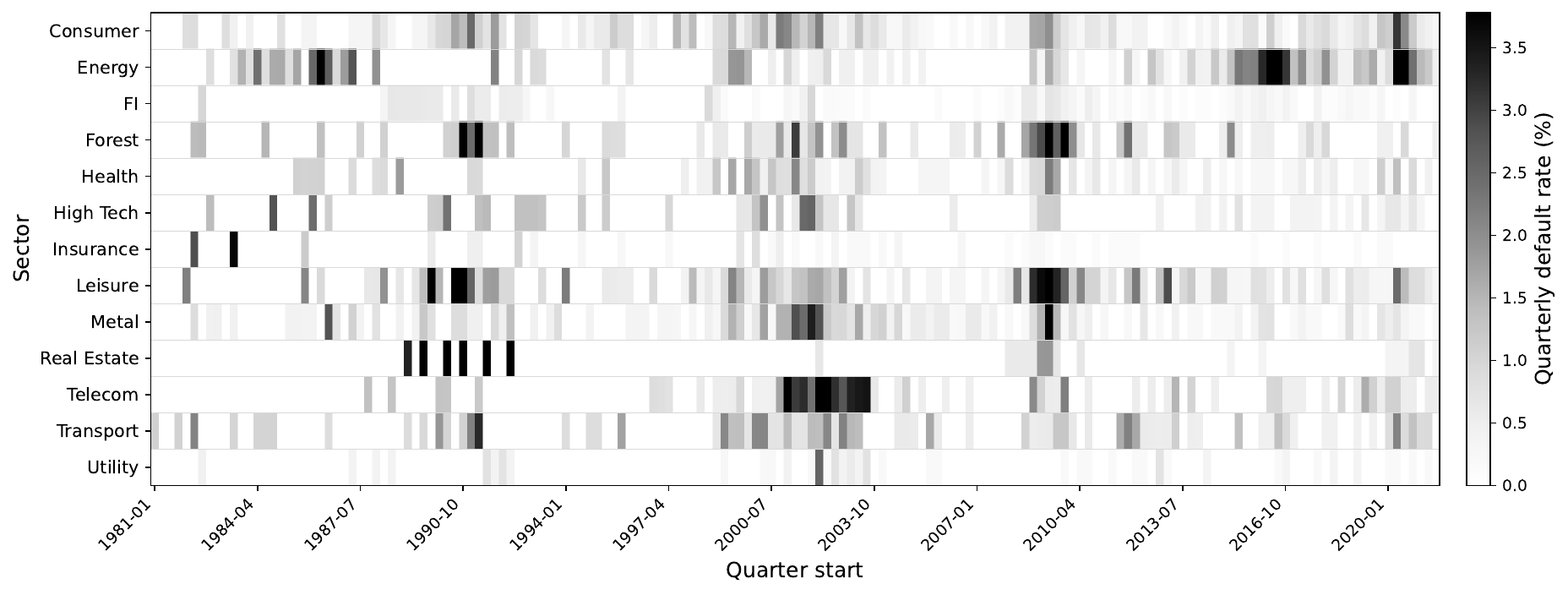}
\caption{
Quarterly sector default-rate heatmap. The underlying panel is monthly, but
three-month aggregation is used only for visualization to reduce sparsity and
improve readability. The color scale reports three-month default rates in
percent. The grayscale intensity is truncated at the 99th percentile of
quarterly default rates to improve the visibility of moderate default-stress
episodes.
}
\label{fig:quarterly_sector_heatmap}
\end{figure}
\end{landscape}

Figure~\ref{fig:quarterly_sector_heatmap} reports a quarterly heatmap of
sector-level default rates. Although the underlying panel is monthly, quarterly
aggregation is used for visualization to reduce sparsity and improve
readability. The main empirical and model-based analyses are conducted at the
monthly scale.

\clearpage

\section{Finite-dimensional representation of temporal coarse-graining}
\label{app:coarse_graining_integrals}

This appendix gives the finite-dimensional integral representation underlying the
temporal coarse-graining construction in
Section~\ref{subsec:temporal_coarse_graining}. The purpose is to make explicit
the mixing distribution, the induced default-count distribution, and the
posterior-implied copula.

For a block of $k$ consecutive months, define
\[
\bm y^{(k)}=(\bm y_1^\top,\ldots,\bm y_k^\top)^\top \in \mathbb{R}^{kS}.
\]
Conditional on the model parameters, the latent block is Gaussian,
\[
\bm y^{(k)} \sim \Normal(\bm\mu^{(k)},\Sigma_y^{(k)}),
\]
where
\[
\bm\mu^{(k)}=(\bm\mu^\top,\ldots,\bm\mu^\top)^\top, \qquad
\left[\Sigma_y^{(k)}\right]_{ij}
=\Omega_{|i-j|},\qquad
i,j=1,\ldots,k.
\]
Here $\Omega_h$ is the lag-$h$ covariance matrix of the monthly probit-scale
sectoral state, including the estimated factor innovation scales and the common
residual scale defined in Section~\ref{subsec:observation_low_rank_state}.

The survival-aggregation map is
\[
T_k(\bm y^{(k)})=\left(
T_{k,1}(\bm y^{(k)}),\ldots,T_{k,S}(\bm y^{(k)})
\right)^\top,
\]
with
\[
T_{k,s}(\bm y^{(k)})=1-\prod_{j=1}^{k}\{1-\Phi(y_{j,s})\},
\qquad s=1,\ldots,S.
\]
Thus the induced $k$-month sectoral default-probability vector is
\[
\bm p^{(k)}=T_k(\bm y^{(k)}).
\]

Let $G_k$ denote the probability law of $\bm p^{(k)}$:
\[
G_k=\mathcal{L}\left(\bm p^{(k)}\right)=
\mathcal{L}\left(T_k(\bm y^{(k)})\right),
\qquad
\bm y^{(k)}
\sim
\Normal(\bm\mu^{(k)},\Sigma_y^{(k)}).
\]
Equivalently, for any measurable set $A\subset[0,1]^S$,
\[
G_k(A)=\int_{\mathbb{R}^{kS}}
\mathbf{1}\{T_k(\bm z)\in A\}
\varphi_{kS}(\bm z;\bm\mu^{(k)},\Sigma_y^{(k)})
\,d\bm z,
\]
where $\varphi_{kS}$ is the $kS$-variate normal density. This representation
shows that $G_k$ is a finite-dimensional Gaussian integral transformed by the
nonlinear survival-aggregation map.

Conditional on $\bm p^{(k)}=\bm p$, sectoral default counts are modeled as
conditionally independent binomial variables,
\[
L_s^{(k)}\mid p_s^{(k)}=p_s,N_s^{(k)}\sim \Bin(N_s^{(k)},p_s),
\qquad s=1,\ldots,S.
\]
Therefore the unconditional $k$-month sectoral default-count distribution is the
multivariate binomial mixture
\[
\Prob(\bm L^{(k)}=\bm\ell)=\int_{[0,1]^S}\prod_{s=1}^{S}
\binom{N_s^{(k)}}{\ell_s}p_s^{\ell_s}(1-p_s)^{N_s^{(k)}-\ell_s}\,dG_k(\bm p).
\]
Hence sectoral dependence at horizon $k$ is generated by the joint mixing
distribution $G_k$, not by conditional dependence in the binomial observation
layer.

The same distribution $G_k$ also defines a copula for the coarse-grained
sectoral default-probability vector. Let $G_{k,s}$ denote the marginal
distribution of $p_s^{(k)}$ under $G_k$, and define
\[
u_s^{(k)}=G_{k,s}(p_s^{(k)}),\qquad s=1,\ldots,S.
\]
The horizon-$k$ copula induced by the dynamic low-rank model is
\[
C_k(u_1,\ldots,u_S)=\Prob\left(u_1^{(k)}\le u_1,\ldots,u_S^{(k)}\le u_S\right).
\]
In the empirical implementation, $G_k$ and $C_k$ are approximated by posterior
samples of the coarse-grained probability vectors
$\bm p_b^{(k,m)}$. The resulting posterior-implied copula is therefore a
sample-based approximation to the copula induced by the Gaussian AR(1) latent
state and the nonlinear survival-aggregation map.

In general, neither $G_k$ nor $C_k$ has an elementary closed-form density,
because $T_k$ contains Gaussian cumulative distribution functions and products
of monthly survival probabilities. Nevertheless, both are analytically
well-defined finite-dimensional objects. This is the key distinction from a
static copula fitted directly at horizon $k$: in the present model, the
horizon-dependent copula is induced by temporally coarse-graining monthly latent
credit-state dynamics.

\clearpage

\section{Bayesian implementation and posterior predictive diagnostics}
\label{app:bayesian_details}

The model is estimated in a Bayesian state-space framework. This appendix
summarizes the prior specification, posterior sampling settings, and posterior
predictive diagnostics used for the fixed-eigenmode models. The same
implementation is used for the factor-selection exercise with
$R=1,2,3,4$, and the $R=2$ specification is used as the main model in the text.

For a model with $R$ fixed eigenmodes, the latent Gaussian score for sector $s$
at month $t$ is
\begin{equation}
y_{t,s}=\mu_s+\sum_{r=1}^R \lambda_{s,r} F_{r,t}+\varepsilon_{t,s},
\end{equation}
where the loading directions $\lambda_{s,r}$ are fixed to the first $R$
empirical eigenvectors of the monthly sectoral correlation matrix. The monthly
default probability is
\begin{equation}
p_{t,s}=\Phi(y_{t,s}),
\end{equation}
and the observed default count is modeled as
\begin{equation}
L_{t,s}\mid p_{t,s},N_{t,s}
\sim
\mathrm{Binomial}(N_{t,s},p_{t,s}).
\end{equation}
The sector intercepts are assigned weakly informative Gaussian priors centered
at empirical initial values,
\begin{equation}
\mu_s\sim\mathrm{Normal}(\mu_{s,0},0.35^2).
\end{equation}
Each latent factor follows an AR(1) process,
\begin{equation}
F_{r,t}=\phi_r F_{r,t-1}+
\sigma_{\eta,r}\eta_{r,t},
\qquad
\eta_{r,t}\sim \mathrm{Normal}(0,1),
\end{equation}
with persistence parameter
\begin{equation}
\phi_r
\sim
\mathrm{TruncatedNormal}
\left(\phi_{r,0},\sigma_{\phi,r}^2;0,0.995\right),
\end{equation}
where $\sigma_{\phi,1}=0.08$ for the first factor and
$\sigma_{\phi,r}=0.10$ for $r\geq 2$. The innovation scale is estimated as
\begin{equation}
\sigma_{\eta,r}
\sim
\mathrm{HalfNormal}(0.25).
\end{equation}
The initial distribution of the AR(1) path is set to the corresponding
stationary distribution,
\begin{equation}
F_{r,0}\sim\mathrm{Normal}\left(
0,
\frac{\sigma_{\eta,r}^2}{1-\phi_r^2}
\right).
\end{equation}
Thus, the marginal variance of each latent factor is not fixed to one, but is
implied by the posterior distribution of $\phi_r$ and $\sigma_{\eta,r}$:
\begin{equation}
\operatorname{Var}(F_{r,t})=
\frac{\sigma_{\eta,r}^2}{1-\phi_r^2}.
\end{equation}
In the implementation, each sampled factor path is centered over time to remove
location confounding with the sector intercepts.

The idiosyncratic residual term is specified with a common scale across sectors,
\begin{equation}
\varepsilon_{t,s}\sim
\mathrm{Normal}(0,\sigma_\varepsilon^2),
\qquad
\sigma_\varepsilon\sim
\mathrm{HalfNormal}(0.25).
\end{equation}
This common-residual-scale specification absorbs local sector-month variation
not represented by the persistent low-rank factors, while avoiding the
over-interpretation of weakly identified sector-specific residual variances.

Posterior sampling is performed using the No-U-Turn Sampler. For each
$R=1,2,3,4$ model in the factor-selection exercise, we use four chains, 5000
tuning iterations per chain, and 5000 posterior draws per chain. The target
acceptance probability is set to $0.99$. Convergence diagnostics are computed
from the resulting posterior samples. The random seed is shifted across factor
counts to make the posterior simulations reproducible while keeping the
different $R$ specifications independent.

For each posterior draw $m$ and aggregation horizon $k$, posterior predictive
counts are generated from Eq.~\eqref{eq:ppc_aggregated_count}. The posterior
predictive monthly-equivalent rate is
\begin{equation}
\widetilde r_{b,s}^{(k,m)}=
\frac{\widetilde L_{b,s}^{(k,m)}}{kN_{b,s}^{(k)}}.
\end{equation}
The posterior predictive correlation matrix
$\widetilde C^{(k,m)}$ is computed over blocks $b$ from the vectors
$\widetilde{\bm r}_b^{(k,m)}$. Let
$\widetilde C^{(k)}_{\mathrm{med}}$ denote the elementwise posterior median of
$\widetilde C^{(k,m)}$. This posterior predictive correlation matrix is
compared with the empirical correlation matrix $\widehat C^{(k)}$ in the
correlation diagnostics reported in the main text.

For the factor-selection exercise, we use the normalized eigenvalue-scaling
curves and the eigenvalue-scaling RMSE defined in Section~\ref{sec:model}. The
RMSE is computed over the aggregation horizons
$\mathcal K=\{1,2,3,4,6,12\}$ for the $R=1,2,3,4$ fixed-eigenmode
specifications under the common-residual-scale specification.

As expected, the posterior mean of the common residual scale decreases as more
fixed eigenmodes are included. However, the posterior diagnostics reported in
Table~\ref{tab:mcmc_phi_diagnostics} show that the additional higher-rank
persistence parameters in the $R=3$ and $R=4$ models are more weakly identified.
We therefore do not interpret the reduction in the common residual scale alone
as evidence in favor of a higher-rank persistent specification.

\clearpage

\section{Sector-wise variance decomposition}
\label{app:variance_decomposition}

Table~\ref{tab:variance_decomposition} reports a sector-wise variance
decomposition for the main $R=2$ fixed-eigenmode model under the
common-residual-scale specification. The table compares the empirical variance
of monthly sectoral default rates with posterior predictive variance
components from the persistent low-rank latent part, binomial observation noise,
their total, and the posterior predictive observed default-rate variance.

The posterior predictive observed variance is close to the empirical variance
for many sectors, but substantial under-representation remains in Insurance,
Real Estate, and, to a lesser extent, Utility. High Tech is not severely
under-represented in posterior predictive observed variance, although its
persistent low-rank latent component accounts for only a small fraction of the
empirical variance. These discrepancies indicate that some local
sector-specific variation cannot be fully absorbed by two persistent common
factors and a single common idiosyncratic residual scale. We therefore treat
them as a limitation of the parsimonious two-factor specification rather than as
evidence that additional stable common factors are necessarily required.

\begin{table}[htbp]
\centering
\scriptsize
\caption{
Sector-wise posterior predictive variance decomposition for the main $R=2$
fixed-eigenmode model under the common-residual-scale specification. The columns
report the empirical variance of the monthly default rate, the variance ratios
of the persistent low-rank latent component, binomial observation noise, their
total, and the posterior predictive observed default-rate variance relative to
the empirical variance. The last two columns report the shares of the latent
component and binomial noise in the latent-plus-noise variance.
}
\label{tab:variance_decomposition}
\begin{tabular}{lrrrrrrr}
\toprule
Sector
& Emp. var.
& Lat./Emp.
& Bin./Emp.
& Total/Emp.
& PPC/Emp.
& Lat. share
& Bin. share \\
\midrule
Consumer    & $7.14{\times}10^{-6}$  & 0.643 & 0.659 & 1.308 & 1.296 & 0.492 & 0.504 \\
Energy      & $2.05{\times}10^{-5}$  & 0.257 & 0.684 & 0.941 & 0.934 & 0.273 & 0.727 \\
FI          & $1.72{\times}10^{-6}$  & 0.177 & 0.657 & 0.834 & 0.827 & 0.212 & 0.787 \\
Forest      & $2.52{\times}10^{-5}$  & 0.440 & 0.670 & 1.113 & 1.113 & 0.396 & 0.602 \\
Health      & $6.42{\times}10^{-6}$  & 0.275 & 0.885 & 1.164 & 1.163 & 0.236 & 0.761 \\
High Tech   & $1.08{\times}10^{-5}$  & 0.099 & 0.728 & 0.828 & 0.825 & 0.120 & 0.879 \\
Insurance   & $5.39{\times}10^{-6}$  & 0.010 & 0.196 & 0.206 & \textbf{0.153} & 0.048 & 0.948 \\
Leisure     & $2.19{\times}10^{-5}$  & 0.607 & 0.650 & 1.256 & 1.248 & 0.483 & 0.518 \\
Metal       & $8.90{\times}10^{-6}$  & 0.617 & 0.484 & 1.099 & 1.082 & 0.561 & 0.440 \\
Real Estate & $3.38{\times}10^{-5}$  & 0.017 & 0.393 & 0.411 & \textbf{0.363} & 0.042 & 0.957 \\
Telecom     & $2.03{\times}10^{-5}$  & 0.451 & 0.605 & 1.057 & 1.048 & 0.427 & 0.573 \\
Transport   & $1.33{\times}10^{-5}$  & 0.237 & 0.899 & 1.134 & 1.135 & 0.209 & 0.793 \\
Utility     & $1.45{\times}10^{-6}$  & 0.180 & 0.511 & 0.702 & \textbf{0.674} & 0.256 & 0.728 \\
\bottomrule
\end{tabular}
\end{table}

\clearpage

\section{Additional forecast diagnostics} 
\label{app:forecast_tables} 

\begin{figure}[htbp]
\centering
\includegraphics[width=\textwidth]{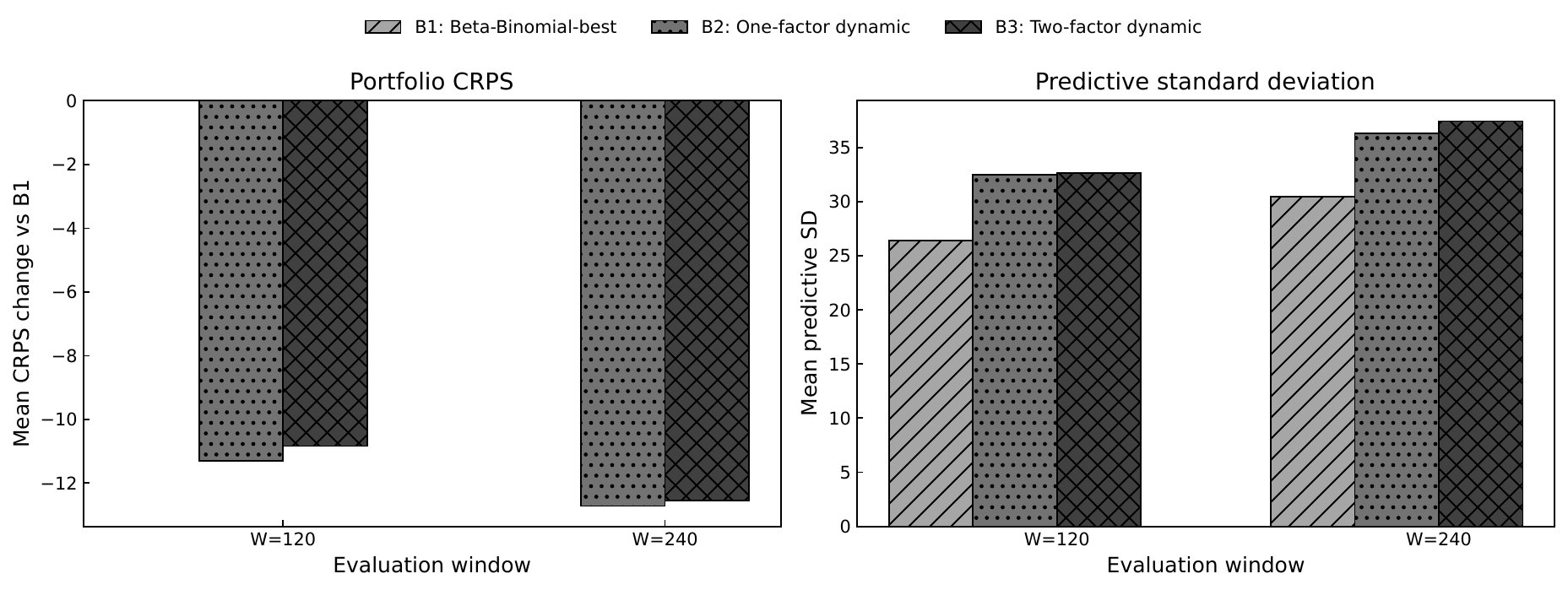}
\caption{Additional annual portfolio forecast diagnostics. The left panel
reports CRPS changes relative to the static beta-binomial baseline B1; negative
values indicate improvement over B1, and B1 is therefore shown at zero. The
right panel reports the predictive standard deviation of the annual total
default count. The static binomial baseline B0 is omitted from this figure
because its predictive dispersion is much smaller and its CRPS is much worse
than those of B1--B3.}
\label{fig:forecast_crps_predsd_appendix}
\end{figure}

This appendix reports the full numerical forecast comparison corresponding to the forecast summary in Section~\ref{sec:forecast}. Table~\ref{tab:forecast_comparison} reports portfolio-level log predictive scores, CRPS, 90\% predictive interval coverage, predictive standard deviations, and sector-vector log scores for all four forecasting models. Figure~\ref{fig:forecast_crps_predsd_appendix} provides additional portfolio-level diagnostics: CRPS changes relative to the static beta-binomial baseline B1 and predictive standard deviations of the annual total default count.

\begin{table}[htbp]
\centering
\caption{
Annual out-of-sample forecast comparison for static and dynamic models. Larger
log scores and sector-vector log scores are better; smaller CRPS is better.
Cov90 is empirical coverage of nominal 90\% predictive intervals. Pred SD is
the predictive standard deviation of the annual total default count. B0 and B1
are favored static baselines with ex-post oracle window selection.
}
\label{tab:forecast_comparison}
\resizebox{\textwidth}{!}{%
\begin{tabular}{llrrrrr}
\toprule
Evaluation window & Model & Log score & CRPS & Cov90 & Pred SD & Sector-vector log score \\
\midrule
$W=120$ & B0: Binomial-best       & -143.089 & 65.634 & 0.000 &  2.350 & -173.714 \\
$W=120$ & B1: Beta-Binomial-best  &   -5.983 & 32.040 & 0.500 & 26.402 &  -40.736 \\
$W=120$ & B2: One-factor dynamic  &   -5.369 & 20.741 & 0.800 & 32.494 &  -37.243 \\
$W=120$ & B3: Two-factor dynamic  &   -5.884 & 21.200 & 0.800 & 32.650 &  -37.693 \\
\midrule
$W=240$ & B0: Binomial-best       & -170.952 & 79.144 & 0.000 &  2.624 & -208.051 \\
$W=240$ & B1: Beta-Binomial-best  &   -6.066 & 36.255 & 0.600 & 30.482 &  -48.803 \\
$W=240$ & B2: One-factor dynamic  &   -5.798 & 23.533 & 0.750 & 36.306 &  -40.385 \\
$W=240$ & B3: Two-factor dynamic  &   -5.042 & 23.698 & 0.900 & 37.442 &  -41.463 \\
\bottomrule
\end{tabular}}
\end{table}

Table~\ref{tab:sector_vector_forecast} reports the corresponding sector-vector
and per-sector forecast diagnostics. The dynamic models substantially improve
over the static baselines in all sector-vector and per-sector measures. In
log-score-based comparisons, the one-factor dynamic model B2 gives the best
sector-vector log score in both evaluation windows, while the two-factor model
B3 is close to B2. By contrast, B3 gives the smallest mean per-sector CRPS in
both evaluation windows. These results indicate that the second factor improves
some aspects of sectoral forecast calibration, but it does not uniformly
dominate the one-factor model in sector-vector predictive density.

Figure~\ref{fig:forecast_crps_predsd_appendix} shows that the dynamic models
reduce portfolio CRPS relative to the beta-binomial baseline B1 and generate
larger predictive standard deviations. The one-factor and two-factor dynamic
models are close in portfolio CRPS, with B2 slightly better in both evaluation
windows.

\begin{table}[htbp]
\centering
\caption{
Sector-vector annual out-of-sample forecast comparison. Larger sector-vector
log scores and per-sector log scores are better; smaller per-sector CRPS is
better.
}
\label{tab:sector_vector_forecast}
\resizebox{\textwidth}{!}{%
\begin{tabular}{llrrr}
\toprule
Evaluation window & Model & Sector-vector log score & Per-sector log score & Per-sector CRPS \\
\midrule
$W=120$ & B0: Binomial-best      & -173.714 & -13.363 & 4.969 \\
$W=120$ & B1: Beta-Binomial-best &  -40.736 &  -3.134 & 3.094 \\
$W=120$ & B2: One-factor dynamic &  -37.243 &  -2.865 & 2.710 \\
$W=120$ & B3: Two-factor dynamic &  -37.693 &  -2.899 & 2.683 \\
\midrule
$W=240$ & B0: Binomial-best      & -208.051 & -16.004 & 5.964 \\
$W=240$ & B1: Beta-Binomial-best &  -48.803 &  -3.754 & 3.763 \\
$W=240$ & B2: One-factor dynamic &  -40.385 &  -3.107 & 3.341 \\
$W=240$ & B3: Two-factor dynamic &  -41.463 &  -3.189 & 3.252 \\
\bottomrule
\end{tabular}}
\end{table}

These results suggest that the second factor mainly affects the calibration of
sectoral default allocations rather than uniformly improving log-score-based
predictive density.

\subsection{Bayesian setup for the rolling forecast experiment}
\label{app:forecast_bayesian_setup}

This subsection summarizes the Bayesian implementation used for the rolling
12-month-ahead forecast experiment in Section~\ref{sec:forecast}. The setup is
the rolling-window analogue of the fixed-eigenmode state-space model used in
the in-sample diagnostics. All quantities used for forecasting are re-estimated
separately at each forecast origin using only the corresponding training
window.

For each training window length $W\in\{120,240\}$ and each forecast origin, we
first construct continuity-corrected monthly probit default rates from the
training sample. The sectoral correlation matrix is computed from these
training-window probit rates, and its leading empirical eigenvectors are used as
fixed loading directions. Thus, the fixed eigenmode loadings used in the
forecasting models are not taken from the full sample. They are recomputed
inside each rolling training window. No observations from the forecast horizon
or from later months enter the construction of the loading directions.

For a model with $R\in\{1,2\}$ factors, the training-window model is
\begin{equation}
y_{t,s}=\mu_s+\sum_{r=1}^{R}\lambda_{s,r}F_{r,t}+\varepsilon_{t,s},
\qquad
p_{t,s}=\Phi(y_{t,s}),
\end{equation}
with the binomial observation equation
\begin{equation}
L_{t,s}\mid p_{t,s},N_{t,s}
\sim
\mathrm{Binomial}(N_{t,s},p_{t,s}).
\end{equation}
The forecasting implementation uses the leading empirical eigenvectors of the
training-window sectoral correlation matrix as fixed loading directions:
\begin{equation}
\lambda_{s,r}=v_{s,r},
\end{equation}
where $v_r$ denotes the $r$th eigenvector computed from the training-window
sectoral correlation matrix. These eigenvectors are used as loading directions
only. No eigenvalue-based amplitude normalization is imposed. The factor
amplitudes are learned through the posterior distribution of the AR(1)
innovation scales, as in the main in-sample specification.

The sector intercepts are assigned Gaussian priors centered at
training-window empirical estimates:
\begin{equation}
\mu_s \sim \mathrm{Normal}(\mu_{s,0},0.50^2).
\end{equation}
Here $\mu_{s,0}$ is obtained from the exposure-weighted training-window default
rate after clipping to avoid boundary probabilities. The AR(1) persistence
parameter of each factor is assigned a truncated Gaussian prior,
\begin{equation}
\phi_r
\sim
\mathrm{TruncatedNormal}
\left(\phi_{r,0},0.12^2;0,0.995\right),
\end{equation}
where $\phi_{r,0}$ is estimated from the training-window PCA factor score and
then clipped to the interval $[0.20,0.98]$ for prior centering.

Each common factor follows a stationary AR(1) process,
\begin{equation}
F_{r,t+1}=\phi_r F_{r,t}+\sigma_{\eta,r}\eta_{r,t+1},
\qquad
\eta_{r,t+1}\sim \mathrm{Normal}(0,1),
\end{equation}
with innovation-scale prior
\begin{equation}
\sigma_{\eta,r}\sim \mathrm{HalfNormal}(0.25).
\end{equation}
The initial state is drawn from the stationary distribution,
\begin{equation}
F_{r,0}\sim\mathrm{Normal}\left(0,\frac{\sigma_{\eta,r}^2}{1-\phi_r^2}\right).
\end{equation}
In the implementation, each latent factor path is centered over the training
window to remove location confounding with the sector intercepts.

The forecast model uses the same common-residual-scale specification as the
main common-eps model:
\begin{equation}
\varepsilon_{t,s}\sim\mathrm{Normal}(0,\sigma_\varepsilon^2),\qquad
\sigma_\varepsilon\sim \mathrm{HalfNormal}(0.25).
\end{equation}
The common residual term is included both in the training-window fit and in the
forecast simulation.

Posterior sampling is performed using the No-U-Turn Sampler. For each rolling
fit, we use four chains, 1000 tuning iterations per chain, and 1000 posterior
draws per chain. The target acceptance probability is set to $0.99$. Random
seeds are fixed from the global seed of the analysis to make the rolling
forecast experiment reproducible.

For each forecast origin, posterior draws are propagated forward for
$H=12$ months according to the estimated AR(1) factor dynamics. Monthly
probabilities are then converted into 12-month sectoral default probabilities by
survival aggregation,
\begin{equation}
p_{t:t+H,s}=1-\prod_{h=1}^{H}{1-p_{t+h,s}}.
\end{equation}
Conditional on these probabilities and beginning-of-horizon exposures, forecast
sectoral default counts are generated from the binomial observation layer. For
forecast scoring, 2000 Monte Carlo predictive draws are generated for each
forecast origin. These draws are used to compute portfolio log scores, CRPS,
90\% predictive interval coverage, predictive standard deviations,
sector-vector log scores, mean per-sector log scores, and mean per-sector CRPS.

The static baselines B0 and B1 are evaluated using candidate rolling windows
\[
W\in\{12,24,36,60,120,180,240\}.
\]
B0 is a sector-specific constant-rate binomial model, and B1 is a
sector-specific constant-rate beta-binomial model. The beta-binomial baseline is
parameterized by an intraclass-correlation parameter $\rho$, with
$\rho\in[10^{-10},0.05]$. As described in the main text, the best static window
is selected ex post from the evaluation sample, so these baselines should be
interpreted as favored static benchmarks rather than real-time forecasting
rules.

\clearpage

\section*{Funding}

This work was supported by JSPS KAKENHI under Grant JP26K06955.

\section*{Declaration of generative AI use}

The authors used ChatGPT by OpenAI to assist with manuscript drafting,
English-language editing, and the generation and refinement of data-analysis
code. ChatGPT was not used as an author and was not used to generate or modify
research data or figure images. All analysis code, numerical results, figures,
and scientific interpretations were checked, revised, and validated by the
authors, who take full responsibility for the content of the manuscript.

\section*{Conflicts of interest}

The authors declare that they have no conflicts of interest.

\section*{Author contributions}

S.M. conceived the study, developed the Bayesian modeling framework, performed
the empirical analysis, generated the numerical results and figures, and wrote
the manuscript. M.H. contributed the theoretical renormalization analysis and
the interpretation of persistent factor amplification under temporal
coarse-graining. Both authors reviewed and approved the final manuscript.

\section*{Data and code availability}

The empirical default data analyzed in this paper are derived from proprietary
historical default datasets and cannot be publicly shared. The analysis and
simulation code is available in a public GitHub repository at

\href{https://github.com/shintaromori/multisector-default-coarse-graining}
{\nolinkurl{https://github.com/shintaromori/multisector-default-coarse-graining}}.

The repository contains the code used to generate the figures and tables,
together with independently generated low-fidelity synthetic data with the same
column structure and time frequency, provided only for code execution and
workflow demonstration.

\bibliographystyle{plainnat}
\bibliography{references}

\end{document}